\documentclass[preprint]{aastex}

\slugcomment{DRAFT \today}

\shortauthors{Hawley, Covey, Knapp et al.}
\shorttitle{M/L/T Dwarfs in the SDSS}
\begin{document}

\title{Characterization of M, L and T Dwarfs in the Sloan Digital Sky Survey\altaffilmark{1} 
}
\author{
Suzanne~L.~Hawley\altaffilmark{2},
Kevin~R.~Covey\altaffilmark{2},
Gillian~R.~Knapp\altaffilmark{3},
David~A.~Golimowski\altaffilmark{4},
Xiaohui~Fan\altaffilmark{5},
Scott~F.~Anderson\altaffilmark{2},
James~E.~Gunn\altaffilmark{3},
Hugh~C.~Harris\altaffilmark{6},
\v{Z}eljko~Ivezi\'{c}\altaffilmark{3},
Gary~M.~Long\altaffilmark{4},
Robert~H.~Lupton\altaffilmark{3},
Peregrine~M.~McGehee\altaffilmark{7}, 
Vijay~Narayanan\altaffilmark{3},
Eric~Peng\altaffilmark{4},
David~Schlegel\altaffilmark{3},
Donald~P.~Schneider\altaffilmark{8},
Emily~Y.~Spahn\altaffilmark{4},
Michael~A.~Strauss\altaffilmark{3},
Paula~Szkody\altaffilmark{2},
Zlatan~Tsvetanov\altaffilmark{4},
Lucianne~M.~Walkowicz\altaffilmark{4},
J.~Brinkmann\altaffilmark{9},
Michael~Harvanek\altaffilmark{9},
Gregory~S.~Hennessy\altaffilmark{10},
S.~J.~Kleinman\altaffilmark{9},
Jurek~Krzesinski\altaffilmark{9,11},
Dan~Long\altaffilmark{9},
Eric~H.~Neilsen\altaffilmark{12},
Peter~R.~Newman\altaffilmark{9},
Atsuko~Nitta\altaffilmark{9},
Stephanie~A.~Snedden\altaffilmark{9},
Donald~G.~York\altaffilmark{13}
}
\altaffiltext{1}{Based on observations obtained with the Sloan Digital
 Sky Survey and the Apache Point Observatory 3.5-meter telescope, which are
 owned and operated by the Astrophysical Research Consortium.}
\altaffiltext{2}{University of Washington, Department of
   Astronomy, Box 351580, Seattle, WA 98195.}
\altaffiltext{3}{Princeton University Observatory, Princeton,
   NJ 08544.}
\altaffiltext{4}{Dept. of Physics and Astronomy,
   Johns Hopkins University, 3701 University Drive, Baltimore, MD 21218.}
\altaffiltext{5}{The Insitute for Advanced Study,
   Princeton, NJ 08540.}
\altaffiltext{6}{US Naval Observatory, Flagstaff Station, 
   P.O. Box 1149, Flagstaff, AZ 86002.}
\altaffiltext{7}{Los Alamos National Laboratory, MS H820,
   Los Alamos, NM 87545.}
\altaffiltext{8}{Dept. of Astronomy and Astrophysics, 
   Pennsylvania State University, University Park, PA 16802.}
\altaffiltext{9}{Apache Point Observatory, P.O. Box 59, 
   Sunspot, NM 88349.}
\altaffiltext{10}{US Naval Observatory, 3450 Massachusetts
   Avenue NW, Washington, DC 20392-5420.}
\altaffiltext{11}{Mt. Suhora Observatory, Cracow 
  Pedagogical University, ul. Podchorazych 2, 30-084 Cracow, Poland.}
\altaffiltext{12}{Fermi National Accelerator Laboratory,
   P.O. Box 500, Batavia, IL 60510.}
\altaffiltext{13}{University of Chicago, Department of Astronomy
   and Astrophysics, 5640 S. Ellis Ave., Chicago, IL 60637.}

Email: slh@astro.washington.edu, covey@astro.washington.edu, gk@astro.princeton.edu, \\
dag@pha.jhu.edu, fan@sns.ias.edu

\section{Abstract}

An extensive sample of M, L and T dwarfs identified in the
Sloan Digital Sky Survey (SDSS) has been compiled.  The sample
of 718 dwarfs includes 677 new objects (629 M dwarfs, 48 L dwarfs) 
together with 41 that have been
previously published.  All new objects and some of the
previously published ones have new optical spectra
obtained either with the SDSS spectrographs or with the
Apache Point Observatory 3.5m ARC telescope.  Spectral types
and SDSS colors are available for all objects; approximately
35\% also have near-infrared magnitudes measured by 2MASS or
on the Mauna Kea system.  We use this sample
to characterize the color--spectral type and color--color relations
of late type dwarfs in the SDSS filters, and to derive spectroscopic
and photometric parallax relations for use in future studies of
the luminosity and mass functions based on SDSS data.  We find
that the $(i^*-z^*)$ and $(i^*-$J) colors provide 
good spectral type and absolute magnitude
(M$_{i^*}$) estimates for M and L dwarfs.
Our distance estimates for the current sample indicate that SDSS is 
finding early M dwarfs out to $\sim$ 1.5 kpc, L dwarfs 
to $\sim$ 100 pc and T dwarfs to $\sim$ 20 pc.  
The T dwarf photometric data show large scatter and
are therefore less reliable for spectral type and distance estimation.

\keywords{stars: late-type; stars: low mass, brown dwarfs 
--- surveys}

\section{Introduction\label{intro}}

Large sky surveys have proven to be
fertile ground for identifying extremely late-type stellar
and substellar objects \citep{nlds}.  
In the past decade, the accumulation of sizable samples of low mass
stars and brown dwarfs has been largely the result of near-infrared surveys
such as the Deep Near-Infrared Survey (DENIS) \citep{d97} and
the Two-Micron All Sky Survey (2MASS) \citep{s97}.
These discoveries culminated in the definition of the L spectral class
\citep{k99, m99, k00, b00}.
Recently, Sloan Digital Sky Survey (SDSS) photometry 
was used to identify the first
isolated field T dwarfs \citep{s99,t00} and later to provide
candidates that filled the gap
separating the L and T spectral sequences \citep{l00}.  Numerous L dwarfs
have also been found from SDSS photometry \citep{f00,s02}.  
The reward of these and additional 2MASS efforts has been a well defined 
spectral sequence from M through T dwarfs \citep{b02,g02}.

The SDSS offers more than just photometric candidates.
The spectral database is expected to ultimately contain more than
one million calibrated spectra, many of which will
be late type dwarfs.
Here we present a study of hundreds of M and L dwarfs compiled from
SDSS spectra, together with spectra of SDSS photometric 
candidates obtained at the 3.5m ARC telescope at Apache Point Observatory 
(APO).  Our final sample also includes all previously published SDSS late-type
dwarfs.  We first describe the assembly of the sample and
the methods used to assign spectral types.  We then investigate 
color--spectral type and color--color relationships, and determine
empirical spectroscopic and photometric parallax relations
in the red SDSS filters.  This paper is intended to provide the
necessary calibration information so that SDSS objects may be
characterized on the basis of their photometry.  Later efforts
will be directed at compiling complete samples of objects and using the
information described here to determine luminosity and mass functions.

\section{Data\label{data}}

Data in our sample include SDSS photometry, 2MASS near-infrared
photometry, and spectra obtained from three distinct sources: 
the SDSS Early Data Release (EDR) \citep{st02}, SDSS spectra taken after the 
EDR, and APO 3.5m telescope followup observations
of SDSS photometric candidates.  The following sections describe these
various data sets.

\subsection {SDSS photometry\label{sdssphot}}

The SDSS mission is to map a
quarter of the sky centered on the North Galactic cap, 
acquiring accurate photometry of 100 million
objects in 5 filters \citep{f96,g98} and
accumulating over 1 million spectra \citep{y00}.
The survey promises to provide astronomers with an unprecedented source of new
astronomical objects.  Both photometric and spectroscopic data are
obtained with a dedicated 2.5 meter telescope at
Apache Point Observatory in New Mexico.  The
telescope's 3 degree field is imaged in the 5 filters ($u$,$g$,$r$,$i$,$z$) 
simultaneously by scanning a great circle across the sky at the sidereal
rate.\footnote{
We adopt italics to denote the filters 
(e.g. $u$), while asterisks indicate the preliminary
calibration of magnitudes within that filter (e.g. $u^*$).}
Photometric data are automatically reduced by SDSS data processing
software
\citep{lup01} and calibrations are obtained from observations with a
20-inch photometric 
telescope at the same site \citep{h01,sm02}.  
\citet{st02} provides information on the central wavelengths and
widths of the SDSS filters.
See the extensive discussions by \citet{l99} and \citet{sm02} 
for further description of the SDSS photometric system.
The late type dwarfs reported here are exceedingly red, and many are not 
detected in the $u$ and $g$ filters.  We discuss only $r^*$, $i^*$ and $z^*$ 
magnitudes in this paper.

There are two sources of uncertainty in the photometric data that
have possible ramifications for late type dwarfs.  First, there are
slight differences in the response of the detectors used for the
$z$ filter observations.  Spectral synthesis of L and T dwarfs
(Burgasser, private communication) indicates that these differences 
may introduce
$z^*$ magnitude errors of up to 0.1 mags for early L through 
late T dwarfs.  We have not attempted to correct for this source
of error.  Second, the magnitudes used here are the observed
magnitudes, with no correction for reddening.  As we show in 
Section~\ref{distances}, some of the early M dwarfs in our sample
lie as much as 1.5 kpc distant.  We examined the reddening
values returned by the SDSS target selection software, which calculates
the total Galactic extinction in the SDSS bands using the maps of
\citet{sch98}.  The reddest objects at each spectral type have 
typical values $\lesssim 0.1$ in $r$, $\lesssim 0.07$ in $i$, and 
$\lesssim 0.05$ in $z$.  These are upper limits, as most objects are 
nearby and will have negligible reddening.  Therefore,
reddening should introduce only a small uncertainty into our 
photometric parallax relations.  

\subsection {SDSS spectroscopy\label{sdssspec}}

Areas of the sky which have been observed by the SDSS photometric
camera provide candidates for spectroscopic observation with
the SDSS twin fiber-fed spectrographs.  The fibers have 3 arcsec diameter
and provide wavelength coverage from 3800\AA\ to 9200\AA\ with 
resolution $\lambda/\Delta\lambda$ $\sim$ 1800.
As described by \citet{st02},
SDSS targets fall into three categories: (1) spectrophotometric 
and extinction calibration stars and sky fibers; (2) galaxies, luminous
red galaxies and quasars, the primary SDSS targets; and (3) additional
objects of interest (various categories of stars, objects of unusual
color) which use excess fibers in fields of low galaxy density.
The last category includes unresolved objects of extremely red color which are
late type dwarf candidates.  Only objects in categories (1) and (2), plus
the very rare brown dwarf candidates, are tiled, i.e. guaranteed a fiber
\citep{st02,bl02},
subject to physical limitations which prevent the placement 
of fibers closer than 55 arcsec apart.  
During the SDSS
commissioning period, the selection criteria for the primary targets
were refined and the efficiency improved.  Because many of the
spectra of late M and L dwarfs are obtained from observations
of quasar candidates, especially high-redshift candidates \citep{r02},
the number of such spectra has been declining
as the survey progresses, though it is still substantial.  

The spectroscopic data are automatically reduced by the
SDSS pipeline software which produces spectrophotometrically
calibrated spectra including telluric correction.  
The pipeline also provides spectral
classification, emission line identification and redshifts.

\subsubsection {Early Data Release spectra\label{edrspec}}

The SDSS EDR contains some 55,000 spectra.
Those spectra that met at least one of the following criteria were
selected as candidate late type dwarfs: 
a) colors of $(r^*-i^*) > 1.8$ and $(i^*-z^*) > 1.0$; 
b) spectrum was identified automatically as a late type star by the
SDSS processing software;  
c) object was targeted for spectral
observation as a possible brown or red dwarf on the basis of photometric
colors.  There is significant overlap between these criteria.  The initial
sample contained $\sim$ 1000 objects, 
which was reduced to 626 objects after removing spectra that were too
noisy to positively identify with a spectral type, or that were clearly
not M, L, or T dwarfs.  No additional spectral reduction or calibration 
was performed aside from rebinning to increase the signal-to-noise ratio.

\begin{figure}
\figurenum{1}
\epsscale{0.95}
\plotone{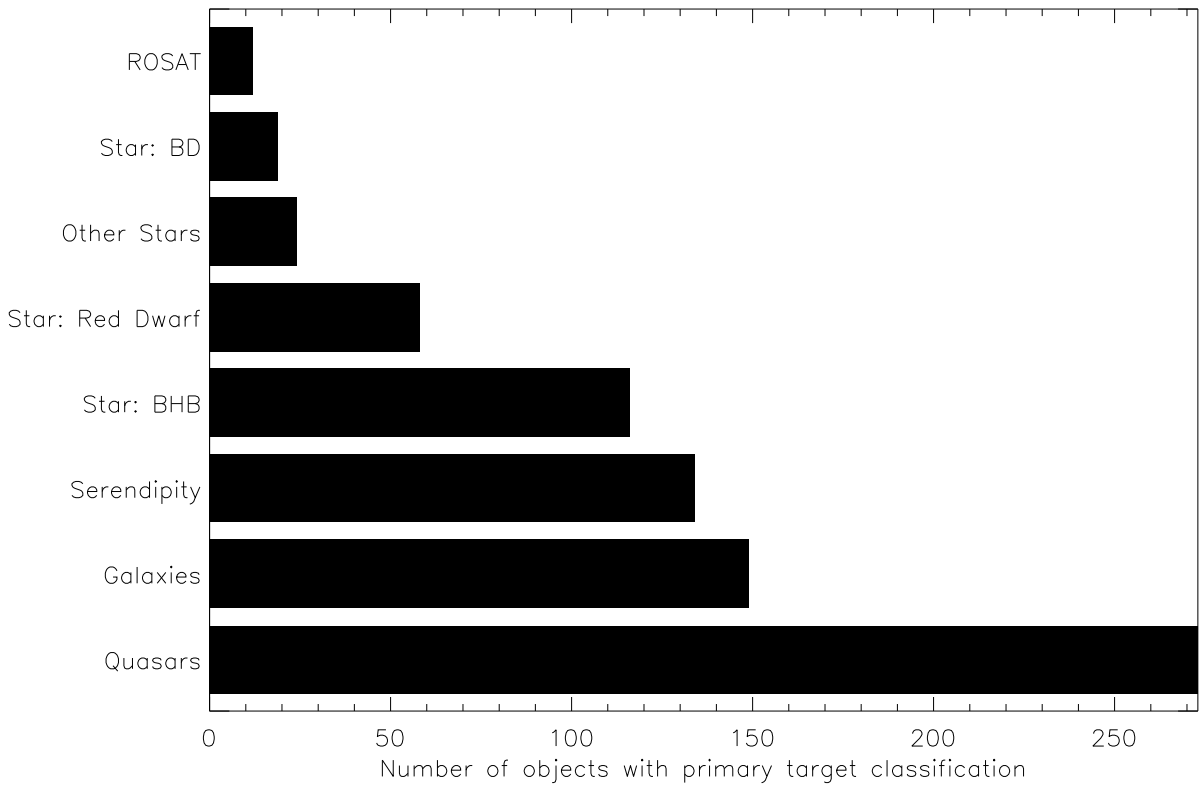}
\caption{SDSS primary targetting assignments are shown for objects that
subsequently turned out to be M and L dwarfs in the SDSS EDR spectra.
Most of the objects were originally targetted as 
quasars and galaxies, or by the serendipity category which targets
objects with unusual colors.  The Star: BD label refers to brown dwarfs.
As described in the text, the large number
of objects in the Star: BHB (blue horizontal branch) category
is the result of a single plate to check objects with stellar colors.}
\label{fig-targetflags}
\end{figure}

Using the EDR sample, we investigated
the behavior and characteristics of the first implementation
of the spectral targetting algorithms. 
Figure~\ref{fig-targetflags} 
shows the distribution by target category for the M and L dwarfs
in our EDR sample.
Clearly, most of the objects were targetted as quasars, galaxies 
or serendipity, the latter because of their unusually red colors.
The categories aimed at discovering red and brown dwarfs
contributed about 15\% of the total, while the other stars and
ROSAT selected targets made up the remainder.
A large number (120) of the 
spectra in our sample, nearly all of M dwarfs, were obtained from a 
single plate of objects targetted to sample the stellar locus and
given the arbitrary designation Star: BHB (blue horizontal branch).
In general, that category does not contribute a significant number
of M and L dwarf targets.

\subsubsection {Recent SDSS spectra \label{recentspec}}

An additional sample of candidate late type dwarfs with SDSS spectra
(the ``recent'' sample) was
created by searching the SDSS spectral database as of July,
2001 for spectra of objects with $(i^*-z^*) > 1.4$ in order to
increase the numbers of very late objects (spectral types M9 and
later).  This sample, once culled 
of objects previously identified in the EDR sample, contained
about 100 candidates.  The number was reduced to 33 after removing 
extremely noisy spectra 
and retaining only objects with late M and L spectra.  Notably, one
previously known T dwarf, SDSSJ1254$-$0122 was found in this sample.
Thus, although the T dwarfs are in general very faint in the SDSS
bandpasses, and the SDSS discoveries to date have been confirmed
using follow-up observations on other telescopes, a few may
emerge with SDSS spectra during the course of the survey.

\subsection {APO spectroscopy\label{apospec}}

Many L and T dwarf candidates identified with SDSS photometry
are too faint to be targetted by the SDSS spectrographs, so programs
of followup observations on larger telescopes have been initiated.  Results
from UKIRT, Keck and HET have been previously published 
(see Section~\ref{previous}).
Here we present a collection of 35 previously unpublished 
spectra obtained with the 3.5m ARC telescope at Apache Point 
Observatory.  Targets were chosen from the photometric database
based on their red colors (typically $(i^*-z^*) > 1.6$). 
The red side of the Double Imaging Spectrograph\footnote{DIS was built
by J. Gunn, M. Carr and R. Lupton.  A description is available at
http://www.apo.nmsu.edu/Instruments/DIS/dis.html}
with the low dispersion grating gave usable wavelength coverage from
6000-10000\AA\ with resolution $\sim$ 600.  Exposure
times of 20 minutes at $i^* \sim 19$ and up to an hour at $i^* \sim 20.5$
gave adequate signal-to-noise ratio for spectral typing purposes.
For one run, the medium dispersion grating was used, giving
resolution $\sim$ 1200 and wavelength coverage 7200-9700\AA.
Standard data reduction was carried out with IRAF\footnote{
IRAF is distributed by the National Optical Astronomy Observatories,
which are operated by the Association of Universities for Research
in Astronomy, Inc., under cooperative agreement with the National
Science Foundation.}
including bias subtraction, flat-field correction, automatic extraction
and sky subtraction, and 
wavelength calibration using exposures of HeNeAr arc lamps.  
Flux standards (typically O subdwarfs and white dwarfs)
were observed each 
night, and all data were placed on a relative flux scale.  On many 
occasions there were clouds and/or light losses from the (1.5 arcsecond) 
slit, so the spectra are not uniformly spectrophotometric. 
We estimate that the relative flux calibration, which is required for
our spectral type template fitting procedure (Section~\ref{template}),
has a typical uncertainty of 10\%.  No telluric corrections were
applied to the APO spectra.

\subsection {2MASS near-infrared photometry\label{twomass}}

To supplement the SDSS color information with near-infrared
photometry,
we searched the public 2MASS database available in December, 2001
for all of our new SDSS objects.  Position matches within
2 arcseconds were accepted as detections; more than 95\%
of the matches were within 1 arcsecond.
Approximately 35\% of our SDSS sample was
detected in 2MASS.  This is almost entirely due to the incomplete sky
coverage of the public 2MASS data release that is currently
available; only four of the M9 and later dwarfs given in
Table~\ref{table-finaldata} that are located in the released
area were not detected by 2MASS.  
These four have very faint expected magnitudes, below the
2MASS survey limits.  They are noted individually in the table.
Thus, when the full 2MASS catalog is made available, we anticipate 
that most of our objects will have measured near-infrared colors.  
Unfortunately, none of the objects with spectral types later 
than L4 and only 6 objects with types L2-L4 were matched, so
our near-infrared information is still quite incomplete at the later types.
Note that the 2MASS magnitudes are calibrated on a Vega-based magnitude 
system while the SDSS magnitudes are on the AB$_\nu$ system \citep{f96}.

\subsection {Previously published data\label{previous}}

We also included data from previously published SDSS late type dwarfs
in the sample \citep{s99,f00,t00,l00,g02,l02,s02}.  
This added 24 objects (14 of the objects with new
optical spectra are previously identified SDSS late type dwarfs), many of them 
late L and T dwarfs that
have been confirmed with near-infrared spectroscopy.  
Three objects were found to have been previously identified by other groups:
2MASSJ1707+6439 (type M9) by \citet{g00}; 
DENISpJ1159+0057 (type L0) by \citet{m99}; 
and 2MASSWJ0801+4628 by \citet{k00}.

\begin{figure}
\figurenum{2}
\epsscale{0.95}
\plotone{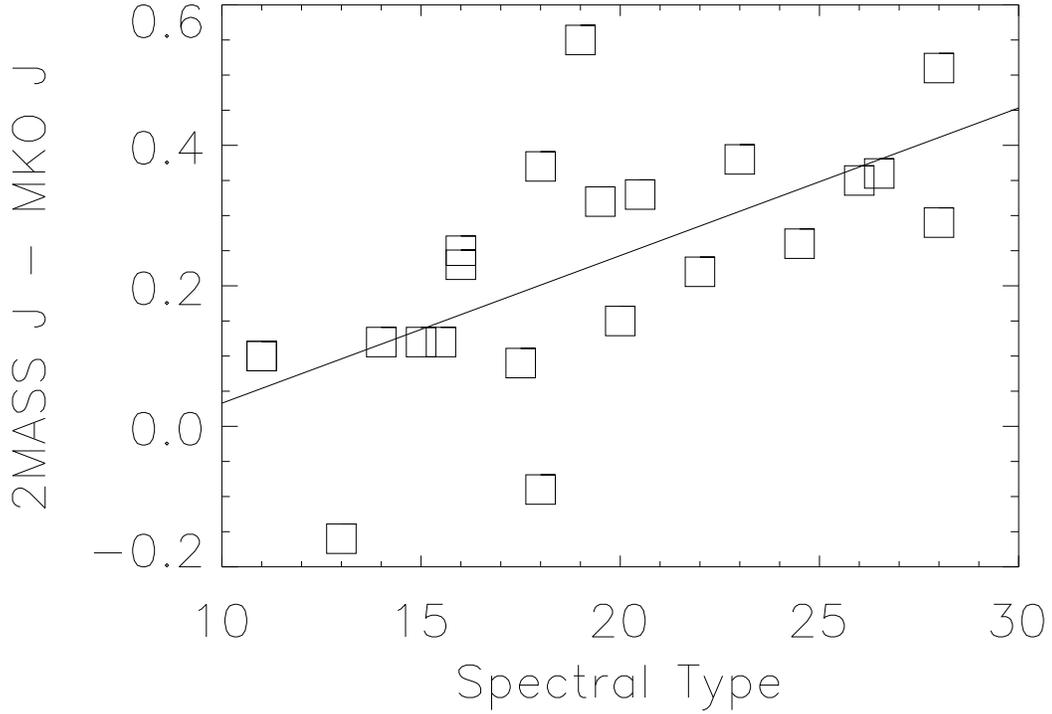}
\caption{The difference between the 2MASS J magnitude and
the MKO J magnitude is plotted for 21 late type dwarfs measured
with both systems.  
Numerical spectral types are assigned as M0=0, M5=5, L0=10,
L5=15, T0=20, T5=25 here and in all subsequent figures where
numerical spectral types are used for fits.  The linear least 
squares fit shown is given in the text.  MKO J magnitudes
given in Table~\ref{table-finaldata}  (and in \citet{l02}
for other previously published objects in the sample),
were transformed to the 2MASS system using this relation.}
\label{fig-mko2mdif}
\end{figure}

Some of the published near-infrared photometry was obtained with
the Mauna Kea Observatory (MKO) system at UKIRT \citep{ha01,l02}.
There are systematic differences
between the MKO and 2MASS photometry for these very cool objects.
Figure~\ref{fig-mko2mdif} shows the difference in measured J band
magnitude as a function of spectral type for 21 objects of type M8 and later
which have been measured on both systems \citep{k99,k00,b02,l02}.  The line
is a simple linear fit:
\begin{equation}
\label{equation-mko2mdif}
$$

J (2MASS) $-$ J (MKO) = $-0.177 + 0.021 \times$ (spectral type)
$$
\end{equation}
applicable between spectral type L0 (10) and T8 (28),
with a dispersion of 0.13 mags.\footnote{
The numerical spectral type in equation~\ref{equation-mko2mdif}
follows standard convention and is used throughout this paper:
M0=0, M5=5, L0=10, L5=15, T0=20, and so forth through T8=28.}
We will use the 2MASS J magnitudes
extensively in our analysis, so for consistency we convert the published
MKO J magnitudes to the 2MASS system using equation~\ref{equation-mko2mdif}.  

The large dispersion present in the measurements is probably
due to the sizable 2MASS uncertainties at these faint magnitudes.
One of us (HCH) has independently investigated the J
magnitude differences between the two systems.  He finds a
similar $\Delta$J relation, indicating that 
equation~\ref{equation-mko2mdif}
is probably more reliable than suggested
by the large scatter, at least among the L dwarfs.  For the T dwarfs,
the data are still too poor to make any firm statement about whether
the dispersion is real.

\section{Spectral typing\label{sptype}}

To assign spectral types to the objects in our sample we first compiled
a set of high signal-to-noise ratio spectral type standards.
We then employed two automatic fitting methods which made use of
complementary information.  In the first method, spectral
indices of individual features were measured and compared to 
a standard sequence, following \citet{rhg95} and \citet{k99} (hereafter K99).
The second method used a least squares fitting algorithm to match primarily 
the continuum shape to the spectral type standards, as described
by \citet{k91} and \citet{h02}. The two methods provided independent 
measures of the 
spectral type.  Each spectrum was then examined visually by two of
us (SLH, KRC) and a final type assigned.

\subsection{Standards\label{stds}}

The spectral type standards used 
as templates in the least squares fitting routine are listed
in Table~\ref{table-templates}.  The late M and L dwarfs were chosen 
from the standards described in K99 and \citet{k00} (hereafter K00), 
and used to define their
L dwarf sequence.  The early-mid M dwarfs
were chosen from the list 
of primary spectral type standards used to define the 
M dwarf sequence by \citet{k91} (except GJ1141B which
was taken from \citet{rhg95}).  
The APO 3.5m data
in Table~\ref{table-templates}
were obtained by us for this purpose, and have the same resolution
and wavelength coverage as the majority of the data 
described in Section~\ref{apospec}. 
The Keck data have been made public courtesy of Neill 
Reid\footnote{See the website
at www.physics.upenn.edu/$\sim$inr for links to these data.}
and are described in K99 and \citet{k00}.

\subsection{Spectral indices method\label{indices}}

The standards listed in Table~\ref{table-templates},
and a number 
of additional objects with well-known spectral types from
the same references, were measured 
to develop polynomial relationships between spectral type and
individual spectral features.
We first implemented the spectral indices suggested in K99
for the L dwarf sequence, which combine measurements of molecular
bands (CrH, TiO, VO) and atomic lines (Rb I, Cs I) as described
therein.  Specifically, we computed their indices CrH-a, Rb-b/TiO-b, 
Cs-a/VO-b and Color-d.  Our task was immediately complicated by
two problems: our spectra have lower resolution and generally
lower signal-to-noise ratio compared to the Keck spectra used by
K99.  Thus, for example, our Rb I and Cs I line measurements were usually
too noisy to give useful results.  We therefore tried numerous
additional diagnostics, settling finally on the set
described in Table~\ref{table-indices}.  In the early-mid
M dwarf range, we also computed the TiO5 and CaH3 indices,
which are defined and used extensively for spectral typing in 
the PMSU survey of 2000 M dwarfs in the Solar Neighborhood
\citep{rhg95,hgr96}.

\begin{figure}
\figurenum{3}
\epsscale{0.95}
\plotone{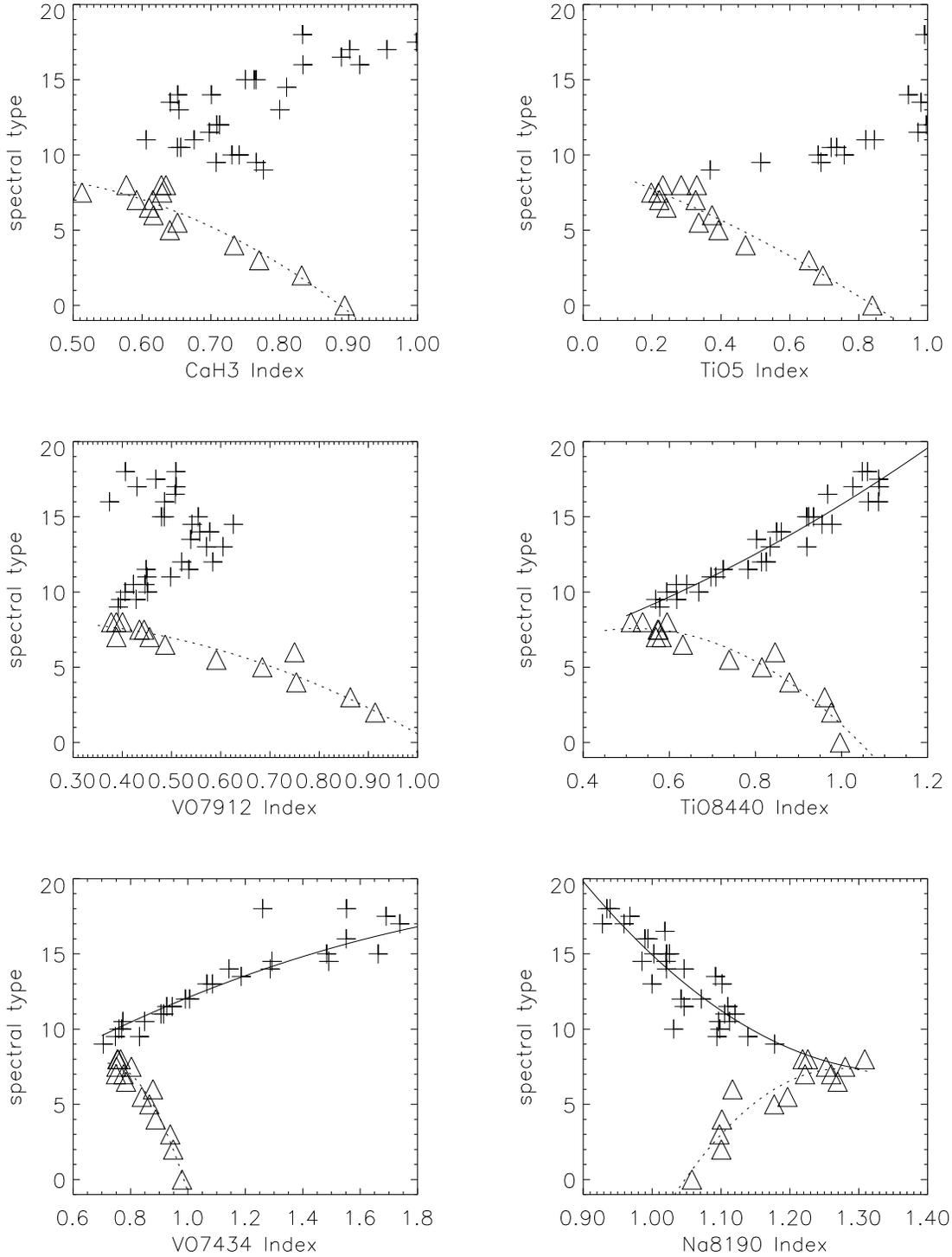}
\caption{Numerical spectral type is plotted vs. the value of the 
spectral index for six different spectral indices.  Open triangles depict M0-M8 dwarfs, while plus signs indicate M9 
and later type dwarfs.  The lines are second-order polynomial, 
least squares fits, with the dotted lines applicable for types M0-M8
and the solid lines applicable for types M9-L8.  The VO7434
fit applies only through type L6 as described in the text.}
\label{fig-indices}
\end{figure}

Figure~\ref{fig-indices} shows most of the new
indices defined here, together with the TiO5 and CaH3 indices
for reference.  Open triangles indicate stars with spectral types M0-M8, 
while the pluses are M9-L8 dwarfs.  All of the indices reverse strength 
near spectral type M8, necessitating two part fits as shown
with dotted and solid lines for the earlier and later types, respectively.
VO7912 reverses strength again near type L4 (similar to the behavior of
CrH-a in the K99 scheme) so we did not use it for fitting the later
type dwarfs.  VO7434 was also useful only to type L6.  Color-d as defined in 
K99 provides useful information, but it includes measurements at red wavelengths
that are not observed in the SDSS spectra.  Therefore we tried
our index Color-1 (not shown in the figure), but it provided 
only limited information on the 
mid-late L dwarfs and was not useful for earlier type dwarfs.
Figure~\ref{fig-indices} highlights the particular difficulty with classifying 
objects in the M7-L1 range as nearly all of the indices overlap in this 
spectral type region.  We relied more heavily on the template fitting results 
and examination by eye for those objects.

The final type produced by the spectral indices method consisted of
an average of the types given by each individual index.  The M and
L types were computed separately for each object using the M and L
fits to the indices, respectively.  The standard deviation of the
average type usually gave a good indication of which type (M or L)
was more appropriate.   However, inspection by eye and comparison
with the template type was required to ensure that the automatic
indices type was reasonable.

\subsection{Template fitting method\label{template}}

Our template fitting method followed closely the procedure outlined by
\citet{h02} (hereafter H02), using a modified version of their ALLSTAR
program.  The SDSS candidate spectrum was 
interpolated onto a 1\AA\ wavelength scale,
smoothed with a 25\AA\ window function, and normalized to the flux
in the region 8651-8661\AA.  This differs slightly from the H02 procedure; 
they do not smooth their spectra, which typically have much higher signal-to-noise
ratio, and they normalize their spectra
at 7500\AA.  We required a location where
there was more flux and hence a better signal-to-noise ratio in the very cool
late M and L dwarfs, and which avoided the broad blue wing of the
K I feature.  As described by H02, to compensate for anomalous 
errors caused by noise in the normalization region, 
the candidate spectrum was multiplied by a factor from 0.9 to 1.1 in 
steps of 0.01.  The template spectra (also normalized to the 8651-8661\AA\ 
region) were then successively fit by least
squares to the 21 scaled candidate spectra, and the best match (lowest
sum of mean squared residuals) for each template was recorded.  Deviant
pixels were treated with 2-$\sigma$ rejection as in the H02 description.
When all templates (M0-L8) had been fit, the 3 best fits were
reported.  We had additional confidence in the fit if the
3 best matches gave both low residuals and neighboring spectral types.

\begin{figure}
\figurenum{4}
\epsscale{0.95}
\plotone{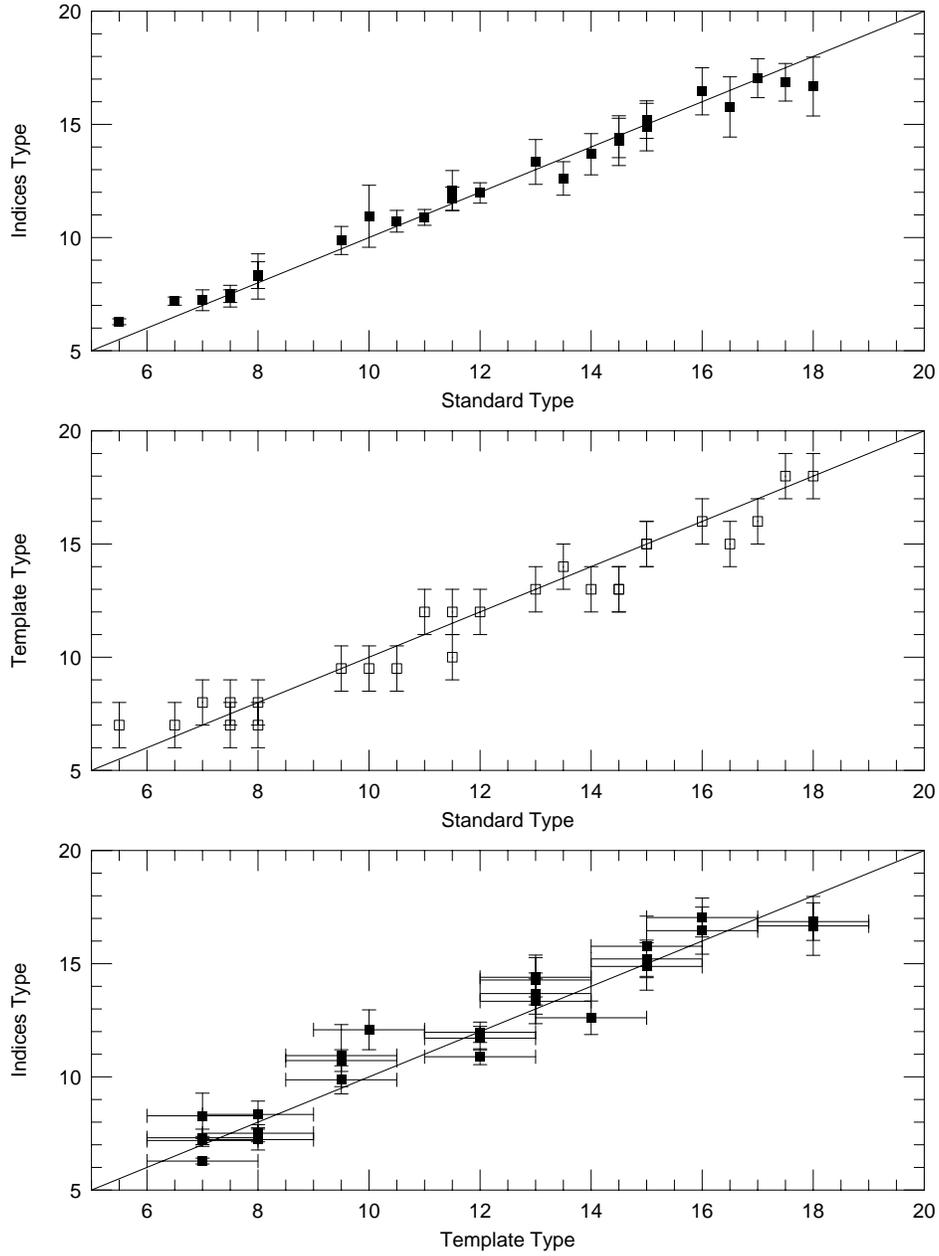}
\caption{The panels show the index and template types compared
with the standard types, and compared with each other.  Only
the additional standards above those given 
in Table~\ref{table-templates} are shown, to avoid matching
templates against themselves.} 
\label{fig-compare}
\end{figure}

Before being implemented on the SDSS sample, the algorithm was first
tested by adding Gaussian noise to spectra of objects with
known spectral types to achieve signal-to-noise ratios 
ranging from 2 to 40 (per \AA), 
and optimizing parameters (smoothing width, normalization location, etc.) such
that the objects were re-identified as correctly as possible, resulting in
the parameters described above.  Our results from testing both
the indices and template methods on spectra of objects with known types
(but not the templates themselves when testing the template method!)
are shown in Figure~\ref{fig-compare}.  The good agreement
indicates that the methods are independently reliable, and
each is able to predict the spectral type within $\sim$
1 spectral class for both M and L dwarfs.

Finally, we note that the template method presumes accurate relative 
flux calibration for both template and program data.  
Both the APO and Keck data, as well as 
the SDSS data, proved to have adequate relative spectrophotometry as shown by 
the good agreement between the template and indices typing methods for
nearly all objects. 

\subsection{Final Types\label{finaltypes}}

After being typed automatically by both the indices and template
methods, the spectra were examined by eye.  If the two measures 
gave different results, a final type was
assigned by visual inspection and comparison to templates.  
There are two previously known T dwarfs for which we have new optical 
spectra.  We classified these as T:: by visual inspection, as the spectra 
appeared to be later than our latest L dwarf template.
Ultimately, 616 M dwarfs, 42 L dwarfs, and 1 T dwarf were typed on the basis
of their SDSS spectra, and an additional 13 Ms, 21 Ls, and 1 T were typed using
APO spectra.  

We estimate that this complete method of spectral typing gives types with
an uncertainty of $\pm$ 1
spectral type for the M and L dwarfs presented here.  
This uncertainty estimate is
bolstered by considering that 1) objects with multiple spectra were 
consistently typed with identical or neighboring spectral types; and 2)
our spectral types match those of other groups within $\sim$ 1 spectral
type for the few stars in common (see Table~\ref{table-finaldata}).  

\section{Results\label{results}}

The large number of early and mid M dwarfs
prohibits showing individual spectra of all objects.  
Instead, we provide the mean and standard deviation for
several optical and near-infrared colors
at each spectral type in Table~\ref{table-summary}.\footnote{We
will make available the data file giving individual information
for each star upon request.  Contact the first author at the 
email address given.}
Our highest signal-to-noise ratio SDSS spectrum for each spectral type 
(M0-L4) is shown in Figure 5(a,b).  The signal-to-noise
ratios range from 10 (at type L4) to 50 (at type M3), measured
at our normalization region near 8650\AA.  This set of
SDSS standard spectra has been submitted to the collaboration
for inclusion in the spectral pipeline software to assist in assigning
automatic types to late type dwarfs during the SDSS processing.\footnote{These
spectra are also available upon request.}

Individual photometric and spectral type information for all dwarfs
with types M9 and later are given in Table~\ref{table-finaldata}.  
Near-infrared data are from 2MASS except where noted, and the K
magnitudes are therefore measured in the K$_s$ filter (see K99
for discussion of this filter bandpass and typical effects
in late type dwarfs).
The M9 and L dwarf spectral types are ours,
with the exception of SDSSJ0830+4828 which has
a previously determined type of L9 based on its near-infrared spectrum 
\citep{g02}.
All objects with previously determined spectral types
are noted, and the references are given in the table notes.  For the
late L dwarfs, differences between optical and
near-infrared spectral types have been
discussed by \citet{g02}.  However, 
the few objects in common in our sample do not appear to differ 
systematically in type.  All T dwarfs have spectral types
determined from near-infrared spectra.

Typical photometric errors for the $r^*$ data in Table~\ref{table-finaldata}
are (in magnitudes): $\lesssim 0.02$ for
$r^* < 20$; 0.05 at $r^* = 21$; 0.1 at $r^* = 22$; 0.25 at $r^* = 23$;
and 0.5 at $r^* = 24$.  As described by \citet{y00} and the table notes,
the nominal 5$\sigma$ detection limit is $r^* = 23.1$.  We have indicated
detections fainter than this limit with an asterisk, and individual
measurements with uncertainty greater than 0.5 magnitudes with a colon.

The $i^*$ and $z^*$ photometry are brighter for these very
red objects.  Photometric errors in $i^*$ range from $\lesssim 0.05$
for $i^* < 20$ to 0.2 at $i^* = 22.5$; corresponding values for $z^*$
are $\lesssim 0.05$ for $z^* < 19$ and 0.2 at $z^* = 21$.  The nominal
5$\sigma$ detection limits are $i^*$ = 22.3 and $z^* = 20.8$.  
Again we indicate detections fainter than these
limits with an asterisk, and note uncertainties greater than 0.2 magnitudes
in the $i^*$ and $z^*$ measurements with a colon.  

Spectra for the M9 and later objects 
are shown in Figure 6(a-f) for the SDSS spectra
and Figure 7(a-d) for the APO spectra.  The SDSS
spectra have been smoothed by 11 pixels 
and the APO spectra
have been smoothed by 3 pixels.  The smoothing
amounts to $\sim$ 20\AA\ in each case.
The signal-to-noise (S/N) ratios range from 2-10 for these
spectra, measured at the normalization region near
8650\AA.  For reference, the M9 dwarf SDSSJ0249-0034
(Figure 6a) has a relatively high S/N ratio $\sim$ 10; the M9 dwarf
SDSSJ1719+6053 (Figure 6b) has moderate S/N ratio $\sim$ 5; and the
L6 dwarf SDSSJ0236+0048 (Figure 6f) has low S/N ratio $\sim$ 2.

\section{Discussion}

The total sample assembled for this paper, including our
new identifications and those previously published, consists of
718 SDSS late type dwarfs (632 M, 76 L, and 10 T). 
All have well-determined spectral types and photometric
measurements in $i^*$ and $z^*$.  Many also have $r^*$ and/or 
near-infrared data.
This sample allows us to investigate the color--spectral type and
color--color relationships in the SDSS and near-infrared colors.
Using a calibrated spectral type -- absolute J magnitude
relation for objects with known distances,
we can determine absolute magnitude relations in the SDSS
filters.  Finally, we use those relations to estimate distances
for the sample.  We caution that the present sample is not complete
within either distance or apparent magnitude limits, as the selection
of spectra from the various SDSS targetting algorithms gives neither
homogeneous nor fully sampled results.  The sample as it stands
is therefore not useful for luminosity or mass function determinations.
Rather, we seek to provide the necessary color, spectral type 
and absolute magnitude calibration information to be used in a 
future study, when complete samples are available.

\subsection{Color -- spectral type relations}

Figure~\ref{fig-avesptcolor} shows the correlation between color 
and spectral type for six different colors:
two optical SDSS colors [$(r^*-i^*), (i^*-z^*)$], three optical--near-infrared
colors [$(i^*-$J), $(z^*-$J), $(z^*-$K)], and one near-infrared color [(J-K)].
We do not present any results for colors using H magnitudes as 
they are very similar to the results for 
colors formed with K magnitudes and do not provide additional useful
diagnostics.
Note again that the
objects with MKO J-band magnitudes have been converted
via equation~\ref{equation-mko2mdif} to the 2MASS J-band system.
No correction was attempted for the MKO K-band magnitudes
so colors that include K photometry may have systematic offsets
for those objects.  We have not used colors that include the K-band
in any of our numerical fitting.

The SDSS $(r^*-i^*)$ color peaks at $\sim$ 2.8 near spectral type M8.  
Later type objects apparently become somewhat bluer, with 
late L dwarfs having similar 
$(r^*-i^*)$ colors to mid M dwarfs.  However,
the $r^*$ photometry is very uncertain for the later objects,
and it may be that ($r^*-i^*$) simply flattens at $\sim$ 2 with large
observational scatter for dwarfs of mid L and later types.
The $(r^*-i^*)$ color by itself is 
not a good diagnostic for dwarfs later than type M8.  

The $(i^*-z^*)$ color does not show such a turnover; it increases
monotonically from early M into the T dwarfs.  However, it is
quite flat at $\sim$ 1.8 between M9-L3 (see also \citet{s02}), 
and appears to flatten
in the T's at $\sim$ 4.0 though with large scatter (probably due
to uncertainty in the $i^*$ measurements at these faint magnitudes).
It is very useful for identifying candidate late type dwarfs:
$(i^*-z^*) > 1.6$ will pick out objects M8 and later (as
noted by \citet{s99} who found
the first isolated T dwarf using its very red $(i^*-z^*)$ color).
If one has only optical (SDSS) colors, it is possible to partially
discriminate for very late objects by first requiring $(i^*-z^*) > 1.6$, and
then choosing objects with bluer $(r^*-i^*)$ colors (e.g. $\lesssim$ 2.0).
A problem with this approach is that those objects
are very faint, and often do not have good $r^*$ photometry, so the
$(r^*-i^*)$ colors may not be reliable.

The (I$-$J) color, where I is the Cousins I band, was shown by \citet{r01}
to be a good spectral type discriminant from M through T.  Here we plot
($i^*-$J), which shows similar behavior to $(i^*-z^*)$, but with the flattening
confined to types L2--L4 at $(i^*-$J) $\sim$ 4.7, and the T dwarfs all having
$(i^*-$J) $\sim$ 7.  An advantage of $(i^*-$J) is that it covers a large total
range in color, varying from 2 $< (i^*-$J) $<$ 8 between early M and late T.

The $(z^*-$J) color is rather flat, with large scatter, through the mid-L
range (L2-L7).  It is not clear if this scatter represents intrinsic
variations in the objects, or if it is just the result of small numbers
and uncertain data.  The $(z^*-$J) color does change systematically through the late-M
and early-L regime, where the spectral types are not easy to
distinguish spectroscopically (see Section~\ref{indices}).  Interestingly,
the $(z^*-$J) color shown here behaves somewhat differently than the (Z-J) color
discussed by \citet{l02}, which uses the MKO Z and J filters.
Their (Z-J) color is flat at $\sim$ 1.7 between mid-L and mid-T types.
Presumably this is related to differences in the filter bandpasses and 
detector sensitivities.

The $(z^*-$K) color is very well behaved from M through mid L types, showing
a monotonic increase with little scatter.  However, it turns over
and becomes blue in the late L and T dwarfs where methane absorption dominates
the K band.

Finally, the (J-K) color has been investigated previously for
late type dwarfs \citep{r93,k99,l02}.  As discussed by those authors,
it is of limited usefulness, showing small range in the M dwarfs
and large scatter among the L dwarfs.  It also 
becomes blue in the T dwarfs, again because of 
methane absorption in the K band.

In summary, $(i^*-z^*)$ and $(i^*-$J) provide spectral type information
across most of the range from M through T.
The $(z^*-$J) and $(z^*-$K) are useful discriminators
in the spectral type ranges where $(i^*-z^*)$
and $(i^*-$J) flatten.  A combination of optical and near-infrared photometry
allows a reasonable determination of the spectral type
at least through the M and L dwarf classes.  The sparse and relatively
uncertain data on T dwarfs make the color-spectral type relations
still poorly defined for those objects.

\subsection{Color -- color relations}

Figure~\ref{fig-colorcolor} shows several color-color relations for 
the sample.  The spectral type of each object is coded by symbol,
given in the figure caption, 
to verify that indeed
the color-color relations are mapping the spectral sequence.
As expected from the behavior of $(r^*-i^*)$ and $(i^*-z^*)$ described
above, the $(r^*-i^*)$ vs $(i^*-z^*)$ diagram is not useful past the mid-M
dwarfs, with objects M5-L4 occupying the same region of
color--color space.  The $(r^*-i^*)$ vs $(i^*-$J) diagram 
is somewhat more spread out,
but still shows significant overlap in spectral type at the same colors.  
Both the $(i^*-z^*,z^*-$J) and the $(i^*-$J,$z^*-$J) diagrams
show significant overlap in the L dwarfs, with the latter
providing a better sequence from M through early L. 
The $(z^*-$J,$z^*-$K) diagram is also well-behaved
in the M and early L regimes, but has significant overlap
between mid and late L dwarfs and turns over in the T dwarfs
where the K band flux is diminished by methane absorption.
Overall, the color--color plot which provides the best
spectral type discrimination is
$(i^*-z^*,i^*-$J) which shows only a mild flattening in the early L's,
but otherwise provides a monotonic sequence from early M through
late T.  The average colors at each spectral type given in
Table~\ref{table-summary} provide the information necessary
to assign a spectral type if optical and near-infrared
photometry are available.

\subsection{Spectroscopic parallaxes\label{specpi}}

Our ultimate goal is to determine photometric relations between
color, spectral type and absolute magnitude that will allow us
to find the spectral type and distance of an object without
needing to take a spectrum.  The previous sections describe
the color-spectral type relations, and we now investigate
parallax relations.  Unfortunately the sample of objects with
both measured trigonometric parallax and measured SDSS colors
is much too small at present to provide a useful calibration.
Instead, we must employ a two-step process where we first use
a well-observed sample of nearby MLT dwarfs with measured trigonometric
parallaxes, 2MASS J magnitudes and spectral types to calibrate a
spectroscopic parallax relation.  We then apply this relation
to our SDSS sample and obtain parallax estimates which we
combine with the SDSS colors to get photometric parallax
relations.  Eventually, as more trigonometric parallaxes are obtained
for SDSS targets, we expect that the SDSS photometric
parallax relation will be calibrated directly and it will not be
necessary to resort to
a spectroscopic parallax relation for calibration.

Figure~\ref{fig-specpical} shows our M$_J$ spectroscopic
parallax relation, incorporating the eight parsec
sample distances \citep{rg97,nlds} for types M0-M7, and distances
quoted by \citet{l02} (see references therein) for
objects with types M8 and later.  New distances from \citet{d02} were 
available for five objects
(DENISpJ0205-1159, 2MASSJ0559-1404, 2MASSJ0850+1057, SDSSJ1254-0122, 
SDSSJ1624+0029).  We applied the
magnitude differences for the resolved L dwarf binaries described by
\citet{r01} to correct the absolute magnitudes for those objects.
Only objects with measured 2MASS J magnitudes are included
(i.e. those objects that were only measured in other systems were
left out rather than being transformed to the 2MASS system).

The early M dwarfs show the well-documented drop in luminosity
near spectral type M4 \citep{hgr96,rg97},
making fitting difficult.  It appears that there is also
a change in slope near spectral type M8, and another drop in
luminosity near type L6.  The latter effect is poorly defined, and clearly
more parallaxes of objects near this type are required.
In addition, spectral types for the late L
and T dwarfs are based on near-infrared spectra, while 
those for the M and early-mid L dwarfs are based on optical
spectra.  This may introduce systematic effects which could contribute
to the apparent discontinuity at L6.  We therefore caution
that the relation for the L6-T8 dwarfs remains ill-defined at present.

Numerous instances of incorrect 
fitting which propagated into incorrect luminosity and mass function
determinations exist in the literature (see \citep{nlds}).
Therefore, we chose to be very conservative in our fit to 
the M$_J$ -- spectral type relation shown here.  We
adopt a fit consisting of several line segments, together with
a single value for type M4.  Our best fit relations are given by:

\begin{equation}
\label{equation-mjfits}
$$

M$_J$ = $6.46 + 0.26 \times$ (spectral type) \ \ \ \ \ \ \ \ \ \ \ \ \ \  (K7-M3)

M$_J$ = 8.34   \ \ \ \ \ \ \ \ \ \ \ \ \ \ \ \ \ \ \ \ \ \ \ \ \ \ \ \ \ \ \ \ (M4)

M$_J$ = $5.73 + 0.74 \times$ (spectral type) \ \ \ \ \ \ \ \  (M5-M7)

M$_J$ = $8.83 + 0.29 \times$ (spectral type) \ \ \ \ \ \ \ \ \ \  (M8-L5)

M$_J$ = $12.13 + 0.14 \times$ (spectral type) \ \ \ \ \ \ \ \  (L6-T8)
$$
\end{equation}

Note that all T dwarfs with known distances are shown, but 
2MASSJ0559-1404 (T4) and Gl 229B (T6), indicated by open
triangles in the figure, are not included in the fit.
The former is omitted because it lies several magnitudes above
the best fit line and has an undue influence on it.  \citet{b01}
discusses this T4 dwarf in detail, and notes that it has an
unusual flux distribution which may indicate it is a binary
system.  Gl 229B has a UKIRT J magnitude and we chose to use 
only those objects with 2MASS J magnitudes.  In any case, it
does not significantly alter the fit.

Figure~\ref{fig-specpical} allows us to assign M$_J$ values to the 
objects with measured J magnitudes in our sample.  With our measured 
colors, we may then compute M$_{r^*}$, M$_{i^*}$ and M$_{z^*}$
spectroscopic parallax relations as shown in Figure~\ref{fig-specpi3}.
The M$_{z^*}$ relation retains some signature of the discontinuity near L6, 
but all three relations are fairly well-behaved.  Absolute magnitudes
may thus be assigned in any of the red SDSS filters if
the spectral type is known.  We have not attempted to fit the other
absolute magnitude - spectral type relations, but 
Table~\ref{table-summary} includes
the M$_J$ values together with the average colors, so
the average absolute magnitude in $r^*$,$i^*$ and $z^*$ may
easily be computed at each spectral type.

\subsection{Photometric parallaxes}

Using the spectroscopic parallax results from the previous section,
we can now investigate photometric parallax relations. 
Figure~\ref{fig-photpiall} shows the results for the colors
$(r^*-i^*), (i^*-z^*), (i^*-$J) and $(z^*-$J).  We did not compute relations 
using colors involving the K band, because those colors turn over and
become bluer in the T dwarfs.  The $(r^*-i^*)$ color is not useful,
as expected from its behavior in Figure~\ref{fig-avesptcolor}.  
The $(z^*-$J) plots show the effect of the discontinuity at L6 quite
strongly, with significant color overlap between early and late L
dwarfs.  Not surprisingly, the best relations are those using 
the $(i^*-z^*)$ and $(i^*-$J) colors.  Figure~\ref{fig-absmags} shows the two
relations for which reasonable
least squares fits were obtained with second order polynomials
to the average colors from Table~\ref{table-summary}.  
The fits are:

\begin{equation}
\label{equation-photfits}
$$

M$_{i^*} = 4.471 + 7.907(i^*-z^*) - 0.837(i^*-z^*)^2$

M$_{i^*} = 1.213 + 4.379(i^*-$J$) - 0.195(i^*-$J)$^2$
$$
\end{equation}

The second fit (M$_{i^*}$, $i^*-$J) has slight systematic residuals
at the discontinuities near M4 $(i^*-$J $\sim$ 3) and L6 $(i^*-$J $\sim$ 5), 
and should be used with caution.

\subsection{Distance distribution\label{distances}}

The approximate distance limits to which the SDSS is probing
the late type dwarf population can be estimated using these parallax
relations.  Figure~\ref{fig-distlog} is a plot of log distance vs spectral
type for all of the sample objects.  The distances were obtained
from the mean spectral type - absolute magnitude relations given
in Table~\ref{table-summary}.  Distances obtained from 
the photometric parallax relations in equation~\ref{equation-photfits}
typically agree within 25\%, and the photometric distance estimates 
are unbiased when compared to those
found from the spectroscopic parallax relations (i.e. the mean
distance differences for the sample are close to zero).

It appears that early M dwarfs are present in the sample out to almost
1.5 kpc, but the distance limit rapidly decreases at later types.
The SDSS imaging has bright magnitude limits of $\sim$ 14
in $r^*$ and $i^*$ and $\sim$ 12 in $z^*$, while spectroscopic targetting is
limited to objects fainter than $r^* \sim 15$ to minimize scattered light
in the spectra \citep{st02}.  This explains the lack of 
nearby early-mid M dwarfs in the sample.  L dwarfs are being observed to 
$\sim$ 100 pc, which is not
unexpected given the magnitude limits ($i^* \sim$ 19-21) of the various
targetting algorithms and the monotonic decrease in luminosity for
objects of later types.  
The T dwarf results imply a distance limit
of $\sim$ 20 pc, but the distance estimates are less reliable 
considering
the large scatter in the calibrated M$_J$ -- spectral type relation
(Figure~\ref{fig-specpical}), the statistical correction which had
to be applied to the measured MKO J magnitudes for several of the
objects, and the sparse sample.  The two SDSS T dwarfs with
trigonometric parallax measurements (SDSSJ1254-0122, 11.8 pc; 
SDSSJ1624+0029, 10.9 pc; \citet{d02}) have predicted distances of 8.9 pc
and 8.1 pc respectively.

\section{Summary}

We have compiled a large sample of M, L and T dwarfs discovered
in the Sloan Digital Sky Survey imaging data.  New objects have
followup SDSS or APO optical spectroscopy.  About 35\% of the sample
also has near-infrared photometry.  In combination with previously
published SDSS objects, a final sample of 718 objects was analyzed.
Color--spectral type and color--color relations for several optical
and near-infrared colors were examined, and the $(i^* - z^*)$ and $(i^*-$J)
colors were found to be the most useful for estimating spectral
types based solely on photometric information.  

We then employed a spectroscopic parallax
relation for nearby dwarfs with known distances to investigate
spectroscopic and photometric parallaxes in the
SDSS filters.  The $(i^*-z^*)$ and $(i^*-$J) colors gave good
estimates of M$_{i^*}$ and hence are useful 
distance indicators.  Our distance estimates show
that SDSS is finding early M dwarfs at distances $>$ 1 kpc,
L dwarfs to distances $\sim$ 100 pc, and T dwarfs out to 20 pc.  
Thus, the SDSS will
provide numerous examples of the latest type dwarfs, and the
early M dwarfs should be useful for probing Galactic structure.
The next step is to compile samples that are complete to some
distance and/or apparent magnitude limit, in order to investigate
the luminosity and mass functions of these late type dwarfs.
Additional efforts will be aimed at investigating the magnetic
activity properties of the sample, the incidence of subdwarfs,
and possible photometric metallicity indicators.

\acknowledgments

The Sloan Digital Sky Survey (SDSS) is a joint project of The University 
of Chicago, Fermilab, the Institute for Advanced Study, the Japan
Participation Group, The Johns Hopkins University, the 
Max-Planck-Institute for Astronomy (MPIA), the Max-Planck-Institute for 
Astrophysics (MPA), New Mexico State University, Princeton University, the
United States Naval Observatory, and the University of Washington. Apache
Point Observatory, site of the SDSS telescopes and the 3.5-meter
telescope, is operated by the Astrophysical Research Consortium (ARC). 

Funding for the project has been provided by the Alfred P.
Sloan Foundation, the SDSS member institutions, the National Aeronautics
and Space Administration, the National Science Foundation, the U.S.
Department of Energy, the Japanese Monbukagakusho, and the Max Planck
Society. The SDSS Web site is http://www.sdss.org/. 

We acknowledge generous support from Princeton University and NASA
via grant NAG5-6734.

This research has made use of NASA's Astrophysics Data System Abstract
Service, the SIMBAD database, operated at CDS, Strasbourg, France, and the 
Two Micron All Sky Survey (2MASS) Second Incremental Release Point Source 
Catalog (PSC), a joint project of the University of Massachusetts and the  
Infrared Processing and Analysis Center/California Institute of
Technology, funded by the National Aeronautics and Space Administration
and the National Science Foundation.

\clearpage

%--------------------------BIBLIOGRAPHY---------------------------

\clearpage
\begin{deluxetable}{lllll}
\tablewidth{0pt}
\tablecaption{Template Information\label{table-templates}}
\tablehead{\colhead{Spectral Type} & \colhead{Object Name} & \colhead{Observatory} }
\startdata
M0 & Gliese 270 & APO \\
M1 & GJ 1141B & APO \\
M2 & Gliese 220 & APO  \\
M3 & Gliese 251 & APO \\
M4 & Gliese 213 & APO \\
M5 & Gliese 51 & APO \\
M6 & 2MASS 0435+1537 & KECK \\
M7 & 2MASS 0055+2756 & KECK \\
M8 & 2MASS 1047+4026 & KECK \\
M9 & 2MASS 0251+2521 & KECK \\
L0 & 2MASS 0345+2540 & KECK \\
L1 & 2MASS 1439+1929 & KECK \\
L2 & Kelu-1 & KECK \\
L3 & DENIS 1058$-$1548 & KECK \\
L4 & 2MASS 1506+1321 & KECK \\
L5 & 2MASS 1507$-$1627 & KECK \\
L6 & 2MASS 0103+1935 & KECK \\
L7 & DENIS 0205$-$1159 & KECK \\
L8 & 2MASS 1632+1904 & KECK \\
\enddata
\end{deluxetable}

\begin{deluxetable}{lll}
\tablewidth{0pt}
\tablecaption{Spectral Type Indices for SDSS M and L Dwarfs   \label{table-indices}}
\tablehead{\colhead{index} & \colhead{numerator} & \colhead{denominator}\\
\colhead{name} & \colhead{\AA} & \colhead{\AA}\\
}
\startdata
VO 7434 & 7430-7470 & 7550-7570  \\
VO 7912 & 7900-7980 & 8400-8420  \\
Na 8190 & 8140-8165 & 8173-8210  \\
TiO 8440 & 8440-8470 & 8400-8420  \\
Color-1 & 8900-9100 & 7350-7550  \\
\enddata
\tablecomments{The weighted (by the variance) average of the flux density
F$_\lambda$
in the numerator and denominator wavelength regions are each computed.  
The index consists of the ratio with its associated uncertainty. }
\end{deluxetable}

\begin{deluxetable}{llllll}
\tablewidth{0pt}
\tablecaption{Average Color by Spectral Type\label{table-summary}}
\tablehead{\colhead{Spectral Type} & \colhead{$r^*-i^*$} & \colhead{$i^*-z^*$} & \colhead{$z^*-$J} & \colhead{$i^*-$J} & \colhead{M$_J$}}

\startdata
M0  &  0.91 (0.24) & 0.49  (0.06) & 1.42  (0.50) & 1.93  (0.50) & 6.45 \\ 
M1 &  0.99  (0.28) & 0.50  (0.29) & 1.25  (0.08) & 1.78  (0.10) & 6.72 \\  
M2 &  1.09  (0.26) & 0.62  (0.17) & 1.36  (0.18) & 1.94  (0.20) & 6.98 \\  
M3 &  1.29  (0.32) & 0.73  (0.24) & 1.38  (0.12) & 2.08  (0.13) & 7.24 \\  
M4 &  1.57  (0.33) & 0.87  (0.32) & 1.52  (0.13) & 2.39  (0.17) & 8.34 \\ 
M5 &  1.98  (0.25) & 1.09  (0.13) & 1.65  (0.09) & 2.72  (0.13) & 9.44 \\
M6 &  2.27  (0.19) & 1.27  (0.14) & 1.74  (0.08) & 2.99  (0.19) & 10.18 \\  
M7 &  2.67  (0.22) & 1.52  (0.11) & 1.95  (0.16) & 3.47  (0.24) & 10.92 \\
M8 &  2.82  (0.28) & 1.62  (0.11) & 2.04  (0.12) & 3.72  (0.19) & 11.14 \\
M9 &  2.89  (0.44) & 1.79  (0.20) & 2.23  (0.14) & 4.09  (0.28) & 11.43 \\ 
L0 &  2.64  (0.25) & 1.85  (0.11) & 2.36  (0.14) & 4.23  (0.19) & 11.72 \\ 
L1 &  2.56  (0.27) & 1.92  (0.08) & 2.52  (0.08) & 4.45  (0.08) & 12.00 \\
L2 &  2.50  (0.47) & 1.85  (0.09) & 2.76  (0.36) & 4.64  (0.45) & 12.29 \\ 
L3 &  2.34  (0.39) & 2.00  (0.24) & 2.71  (0.19) & 4.84  (0.50) & 12.58 \\ 
L4 &  2.38  (0.27) & 2.26  (0.13) & 2.80  (0.08) & 5.06  (0.18) & 12.87 \\
L5 &  2.00  (0.40) & 2.47  (0.12) & 2.74  (0.23) & 5.21  (0.32) & 13.16 \\  
L6 &  2.79  (0.69) & 2.40  (0.19) & 2.86  (0.15) & 5.27  (0.20) & 14.31 \\
L7 &  2.31  (0.50) & 2.98  (0.50) & 2.93  (0.50) & 5.91  (0.50) & 14.45 \\  
L8 &  2.19  (0.91) & 3.15  (0.32) & 2.93  (0.02) & 6.11  (0.47) & 14.58 \\
L9 &  2.33  (0.50) & 3.18  (0.50) & 2.86  (0.50) & 6.04  (0.50) & 14.72 \\  
T0 &  2.43  (0.50) & 2.88  (0.50) & 2.88  (0.50) & 5.76  (0.50) & 14.86 \\
T1 &  1.47  (0.50) & 3.39  (0.50) & 3.20  (0.50) & 6.59  (0.50) & 14.99 \\
T2 &  1.80  (0.50) & 4.25  (0.50) & 3.13  (0.50) & 7.38  (0.50) & 15.13 \\
T3 &  1.16  (0.33) & 3.48  (0.56) & 3.27  (0.31) & 6.75  (0.87) & 15.27 \\
T4 &  1.04  (0.52) & 3.82  (2.02) & 3.53  (0.06) & 7.35  (2.09) & 15.40 \\
T5 &  *** & *** & *** & *** & 15.54 \\
T6 &  1.96  (0.96) & 4.00  (0.33) & 3.49  (0.06) & 7.49  (0.33) & 15.68 \\

\enddata
\tablecomments{The mean and ($\sigma$) of each color are given.  If 
there is only
one measurement, $\sigma$ is set equal to (0.50).  The J-band absolute
magnitudes come from the calibrated spectroscopic parallax relation described
in section~\ref{specpi} and given in equation~\ref{equation-mjfits}.}
\end{deluxetable}

\begin{deluxetable}{lllllllllll}
\tablewidth{0pt}
\tabletypesize{\tiny}
\tablecaption{SDSS M9, L and T dwarfs with new optical spectra\label{table-finaldata}}
\tablehead{\colhead{Name} & \colhead{RA} & \colhead{Dec} &
\colhead{$r^*$} &\colhead{$i^*$} & \colhead{$z^*$} & \colhead{J} 
& \colhead{H} & \colhead{K$^{12}$} & \colhead{Spectral} & \colhead{Source}\\
\colhead{} & \colhead{J2000} &\colhead{J2000} & \colhead{} &\colhead{} & 
\colhead{} & \colhead{} & \colhead{} & \colhead{} & \colhead{Type} 
& \colhead{}\\
}
\startdata
SDSSJ0022$-$0110 & 00:22:09.31 & $-$01:10:40.2 & 22.69 & 19.77 & 17.88 & 15.83 & 15.10 & 14.63 & M9 &    EDR \\
SDSSJ0223+0006 & 02:23:57.93 & +00:06:22.4 & 23.16* & 20.49 & 18.72 & 16.56 & 16.01 & 15.57 & M9 &    EDR \\
SDSSJ0227+0026 & 02:27:54.92 & +00:26:20.0 & 23.73*: & 20.55 & 18.81 & ***** & ***** & ***** & M9 &    EDR \\
SDSSJ0228+0040 & 02:28:15.65 & +00:40:26.6 & 23.42* & 20.84 & 19.30 & ***** & ***** & ***** & M9 &    EDR \\
\smallskip
SDSSJ0230$-$0012 & 02:30:44.46 & $-$00:12:59.3 & 24.27*: & 20.64 & 19.18 & 16.99 & 16.44 & 15.19 & M9 &    EDR \\
SDSSJ0233$-$0023 & 02:33:11.06 & $-$00:23:30.7 & 22.72 & 20.26 & 18.84 & ***** & ***** & ***** & M9 &    EDR$^{11}$ \\
SDSSJ0249$-$0034 & 02:49:58.36 & $-$00:34:10.0 & 22.16 & 19.38 & 17.64 & 15.46 & 14.80 & 14.28 & M9 &    EDR \\
SDSSJ0309$-$0753 & 03:09:53.46 & $-$07:53:15.3 & 23.36* & 21.20: & 18.60 & 16.50 & 15.68 & 15.40 & M9 &    APO \\
SDSSJ0443+0002 & 04:43:37.60 & +00:02:05.2 & 19.74 & 16.93 & 15.03 & 12.52 & 11.80 & 11.17 & M9 &    APO \\
\smallskip
SDSSJ1050+0058 & 10:50:12.49 & +00:58:03.3 & 23.71*: & 20.47 & 18.57 & ***** & ***** & ***** & M9 &    EDR \\
SDSSJ1134+0022 & 11:34:54.89 & +00:22:54.5 & 19.57 & 16.82 & 15.05 & ***** & ***** & ***** & M9 &    APO \\
SDSSJ1342$-$0027 & 13:42:28.41 & $-$00:27:05.7 & 22.26 & 19.76 & 17.93 & 15.59 & 14.92 & 14.32 & M9 &    EDR \\
SDSSJ1413+0016 & 14:13:21.49 & +00:16:37.3 & 23.33* & 20.15 & 18.53 & ***** & ***** & ***** & M9 &    EDR \\
SDSSJ1413+5154 & 14:13:33.87 & +51:54:17.2 & 22.85 & 19.78 & 17.94 & ***** & ***** & ***** & M9 &    APO$^{8}$ \\
\smallskip
SDSSJ1440+0051 & 14:40:50.61 & +00:51:52.3 & 23.21* & 20.08 & 18.45 & ***** & ***** & ***** & M9 &    EDR \\
SDSSJ1635+0008 & 16:35:21.33 & +00:08:24.3 & 22.56 & 19.87 & 18.10 & ***** & ***** & ***** & M9 &    APO$^{8}$ \\
SDSSJ1648$-$0012 & 16:48:21.07 & $-$00:12:27.1 & 20.54 & 17.67 & 15.85 & ***** & ***** & ***** & M9 &    APO$^{8}$ \\
SDSSJ1707+6439 & 17:07:18.32 & +64:39:33.2 & 19.39 & 16.59 & 14.72 & 12.56 & 11.83 & 11.39 & M9 (M9)$^9$ &    APO$^{9}$ \\
SDSSJ1714+6351 & 17:14:28.46 & +63:51:30.2 & 23.55*: & 20.49 & 18.88 & ***** & ***** & ***** & M9 &    EDR$^{11}$ \\
\smallskip
SDSSJ1716+5537 & 17:16:12.81 & +55:37:54.6 & 22.82 & 19.77 & 17.85 & ***** & ***** & ***** & M9 &    APO \\
SDSSJ1719+6053 & 17:19:13.79 & +60:53:10.5 & 22.24 & 20.01 & 18.26 & ***** & ***** & ***** & M9 &    EDR \\
SDSSJ1723+6321 & 17:23:17.51 & +63:21:19.2 & 22.82 & 20.09 & 18.25 & ***** & ***** & ***** & M9 &    EDR \\
SDSSJ1734+5355 & 17:34:17.52 & +53:55:20.7 & 22.17 & 19.25 & 17.32 & 15.06 & 14.34 & 13.99 & M9 &    APO \\
SDSSJ1737+5617 & 17:37:49.86 & +56:17:00.1 & 23.71* & 20.56 & 18.92 & ***** & ***** & ***** & M9 &    EDR \\
\smallskip
SDSSJ1742+5611 & 17:42:16.83 & +56:11:08.2 & 24.60*: & 20.46 & 18.83 & ***** & ***** & ***** & M9 &    EDR \\
SDSSJ2147+0101 & 21:47:27.63 & +01:01:04.2 & 21.67 & 18.88 & 17.02 & ***** & ***** & ***** & M9 &    APO \\
SDSSJ2204$-$0036 & 22:04:41.97 & $-$00:36:51.4 & 22.73 & 19.80 & 17.91 & 15.70 & 15.08 & 14.37 & M9 &    APO \\
SDSSJ2230+0111 & 22:30:42.38 & +01:11:50.2 & 22.72 & 19.76 & 17.98 & ***** & ***** & ***** & M9 &    APO \\
SDSSJ2244$-$0029 & 22:44:39.86 & $-$00:29:39.0 & 22.24 & 19.60 & 17.86 & 15.82 & 15.03 & 14.62 & M9 &    APO \\
\smallskip
SDSSJ0035+1447 & 00:35:24.44 & +14:47:39.8 & 23.76*: & 21.10 & 19.33 & ***** & ***** & ***** & L0 & recent$^{11}$ \\
SDSSJ0041+1341 & 00:41:54.54 & +13:41:35.5 & 21.24 & 18.78 & 16.92 & 14.61 & 13.78 & 13.39 & L0 &    APO \\
SDSSJ0227$-$0055 & 02:27:23.77 & $-$00:55:18.6 & 23.27* & 20.94 & 19.37 & ***** & ***** & ***** & L0 &    EDR$^{11}$ \\
SDSSJ0256+0110 & 02:56:01.86 & +01:10:47.2 & 23.56* & 20.47 & 18.69 & ***** & ***** & ***** & L0 &    EDR \\
SDSSJ0357$-$0641 & 03:57:21.11 & $-$06:41:26.0 & 22.92 & 20.18 & 18.26 & 15.97 & 15.04 & 14.60 & L0 & recent \\
\smallskip
SDSSJ0747+3947 & 07:47:56.31 & +39:47:32.9 & 21.81 & 19.22 & 17.42 & 15.12 & 14.19 & 13.71 & L0 &    APO \\
SDSSJ0752+4136 & 07:52:59.43 & +41:36:34.6 & 24.08*: & 21.03 & 18.96 & 16.33 & 15.66 & 15.17 & L0 & recent \\
SDSSJ1113$-$0002 & 11:13:16.95 & $-$00:02:46.6 & 21.91 & 19.43 & 17.50 & 15.08 & 14.29 & 13.77 & L0 &    EDR \\
SDSSJ1138+6740 & 11:38:33.10 & +67:40:40.3 & 21.82 & 19.35 & 17.52 & 15.20 & 14.50 & 13.95 & L0 & recent \\
SDSSJ1148+0254 & 11:48:04.26 & +02:54:05.7 & 22.59 & 20.34 & 18.52 & ***** & ***** & ***** & L0 & recent \\
\smallskip
SDSSJ1159+0057 & 11:59:38.50 & +00:57:26.9 & 21.14 & 18.57 & 16.72 & ***** & ***** & ***** & L0 (L0)$^{10}$&    EDR$^{10}$ \\
SDSSJ1228+0050 & 12:28:55.38 & +00:50:44.1 & 23.09 & 20.17 & 18.18 & ***** & ***** & ***** & L0 &    EDR \\
SDSSJ1430+0013 & 14:30:55.90 & +00:13:52.1 & 23.03 & 20.43 & 18.54 & ***** & ***** & ***** & L0 (M8)$^{6}$& EDR$^{3}$ \\
SDSSJ1435$-$0046 & 14:35:17.20 & $-$00:46:12.9 & 22.99 & 20.41 & 18.54 & ***** & ***** & ***** & L0 &    APO \\
SDSSJ1555+0017 & 15:55:26.15 & +00:17:20.6 & 22.48 & 19.27 & 17.39 & ***** & ***** & ***** & L0 &    APO$^{8}$ \\
\smallskip
SDSSJ1722+6329 & 17:22:44.32 & +63:29:46.8 & 22.47 & 19.74 & 17.86 & ***** & ***** & ***** & L0 &    EDR \\
SDSSJ0019+0030 & 00:19:11.65 & +00:30:17.8 & 21.95 & 19.42 & 17.52 & ***** & ***** & ***** & L1 & recent \\
SDSSJ0038+1343 & 00:38:43.99 & +13:43:39.5 & 23.34* & 20.33 & 18.46 & 15.92 & 15.16 & 14.72 & L1 & recent \\
SDSSJ0042+1459 & 00:42:21.15 & +14:59:23.9 & 23.62*: & 21.05 & 19.06 & 16.62 & 15.86 & 15.56 & L1 &    APO \\
SDSSJ0054$-$0031 & 00:54:06.55 & $-$00:31:01.8 & 22.76 & 20.20 & 18.20 & 15.75 & 14.91 & 14.38 & L1 (L2)$^{6}$ & recent$^{3}$ \\
\smallskip
SDSSJ0211+1410 & 02:11:28.25 & +14:10:03.8 & 23.62* & 20.55 & 18.62 & 16.09 & 15.39 & 14.93 & L1 & recent \\
SDSSJ0350$-$0518 & 03:50:48.62 & $-$05:18:12.8 & 23.43* & 20.79 & 18.91 & 16.31 & 15.60 & 15.08 & L1 & recent \\
SDSSJ0626+0029 & 06:26:21.22 & +00:29:34.2 & 22.91 & 20.43 & 18.42 & ***** & ***** & ***** & L1 &    AP0$^{8}$ \\
SDSSJ0815+4524 & 08:15:56.74 & +45:24:11.8 & 22.76 & 20.60 & 18.54 & 16.02 & 15.24 & 14.81 & L1 & recent \\
SDSSJ0927+6027 & 09:27:57.46 & +60:27:46.3 & 22.69 & 19.91 & 18.07 & ***** & ***** & ***** & L1 & recent \\
\smallskip
SDSSJ1045$-$0149 & 10:45:23.98 & $-$01:49:57.7 & 20.05 & 17.58 & 15.77 & 13.13 & 12.37 & 11.81 & L1 &    APO \\
SDSSJ1048+0111 & 10:48:42.84 & +01:11:58.5 & 19.77 & 17.23 & 15.42 & ***** & ***** & ***** & L1 & recent \\
SDSSJ1148+0203 & 11:48:05.02 & +02:03:50.9 & 22.27 & 19.90 & 17.90 & ***** & ***** & ***** & L1 & recent \\
SDSSJ1402+0148 & 14:02:31.75 & +01:48:30.3 & 22.33 & 19.93 & 17.97 & ***** & ***** & ***** & L1 & recent \\
SDSSJ1404+0235 & 14:04:41.68 & +02:35:50.1 & 22.98 & 20.00 & 18.08 & ***** & ***** & ***** & L1 & recent \\
\smallskip
SDSSJ1439+0317 & 14:39:33.44 & +03:17:59.2 & 23.29* & 20.66 & 18.71 & ***** & ***** & ***** & L1 & recent \\
SDSSJ1502+6138 & 15:02:40.80 & +61:38:15.5 & 23.28* & 20.68 & 18.79 & 16.37 & 16.07 & 15.36 & L1 & recent \\
SDSSJ1630+0051 & 16:30:50.01 & +00:51:01.3 & 22.86 & 20.55 & 18.81 & ***** & ***** & ***** & L1 & recent$^{8}$ \\
SDSSJ1728+5845 & 17:28:22.19 & +58:45:09.9 & 23.24 & 21.04 & 19.05 & ***** & ***** & ***** & L1 & recent \\
SDSSJ0235$-$0849 & 02:35:47.56 & $-$08:49:19.8 & 22.34 & 19.89 & 18.01 & 15.52 & 14.80 & 14.20 & L2 & recent \\
\smallskip
SDSSJ0800+4658 & 08:00:48.13 & +46:58:25.5 & 22.45 & 20.00 & 18.12 & 15.48 & 14.57 & 14.27 & L2 & recent \\
SDSSJ1547+0336 & 15:47:27.23 & +03:36:36.3 & 23.41* & 20.46 & 18.68 & ***** & ***** & ***** & L2 & recent \\
SDSSJ1614+0046 & 16:14:20.50 & +00:46:43.6 & 23.14 & 20.89 & 19.10 & ***** & ***** & ***** & L2 & recent$^{8}$ \\
SDSSJ1619+0050 & 16:19:28.31 & +00:50:11.9 & 21.54 & 18.99 & 17.11 & ***** & ***** & ***** & L2 & recent$^{8}$ \\
SDSSJ2259$-$0051 & 22:59:13.88 & $-$00:51:58.2 & 23.81*: & 20.68 & 18.93 & 16.31 & 15.37 & 14.99 & L2 &    APO \\
\smallskip
SDSSJ0207+1355 & 02:07:35.60 & +13:55:56.3 & 22.34 & 19.86 & 18.07 & 15.42 & 14.43 & 13.79 & L3 & recent \\
SDSSJ0328+0032 & 03:28:17.38 & +00:32:57.2 & 22.90 & 20.61 & 18.88 & ***** & ***** & ***** & L3 (L2.5)$^{6}$ & recent$^{3}$ \\
SDSSJ1435$-$0043 & 14:35:35.72 & $-$00:43:47.0 & 24.30*: & 20.90 & 19.02 & ***** & ***** & ***** & L3 &    APO \\
SDSSJ1440+0026 & 14:40:16.20 & +00:26:38.9 & 22.94 & 20.71 & 18.77 & ***** & ***** & ***** & L3 & recent \\
SDSSJ1653+6231 & 16:53:29.69 & +62:31:36.5 & 21.54 & 19.51 & 17.61 & ***** & ***** & ***** & L3 &    EDR \\
\smallskip
SDSSJ2028+0052 & 20:28:20.32 & +00:52:26.5 & 21.37 & 18.89 & 17.04 & ***** & ***** & ***** & L3 &    APO$^{8}$ \\
SDSSJ2140+0112 & 21:40:46.55 & +01:12:59.7 & 23.38* & 21.10 & 18.96 & ***** & ***** & ***** & L3 &    APO \\
SDSSJ2249+0044 & 22:49:53.46 & +00:44:04.6 & 23.98*: & 22.03: & 19.44 & 16.46$^{7}$ & 15.42$^{7}$ & 14.43$^{7}$ & L3 (L5)$^{5}$ &    APO$^{4}$ \\
SDSSJ0330$-$0025 & 03:30:35.12 & $-$00:25:34.6 & 22.31 & 20.14 & 18.04 & 15.29 & 14.42 & 13.83 & L4 (L2)$^{6}$ & recent$^{1}$ \\
SDSSJ0805+4812 & 08:05:31.84 & +48:12:33.0 & 22.72 & 19.90 & 17.59 & 14.71 & 13.91 & 13.42 & L4 &    APO \\
\smallskip
SDSSJ1257$-$0113 & 12:57:37.26 & $-$01:13:36.1 & 22.94 & 20.78 & 18.55 & 15.86 & 14.70 & 14.13 & L4 (L5)$^{5}$& recent$^{4}$ \\
SDSSJ1717+6526 & 17:17:14.10 & +65:26:22.2 & 22.54 & 20.21 & 17.77 & 14.94 & 13.85 & 13.20 & L4 &    APO \\
SDSSJ0127+1354 & 01:27:43.51 & +13:54:20.9 & 23.57* & 22.08 & 19.61 & ***** & ***** & ***** & L5 &    APO \\
SDSSJ0801+4628 & 08:01:40.53 & +46:28:49.5 & 23.36* & 21.28 & 18.80 & 16.29 & 15.44 & 14.54 & L5 (L6.5)$^{13}$ &    APO$^{13}$ \\
SDSSJ0236+0048 & 02:36:17.94 & +00:48:54.8 & 25.04*: & 21.50 & 18.92 & 16.01$^{7}$ & 15.16$^{7}$ & 14.54$^{7}$ & L6 (L6.5)$^{5}$ (L6)$^{6}$&    EDR$^{3}$ \\
\smallskip
SDSSJ1331$-$0116 & 13:31:48.92 & $-$01:16:51.4 & 22.78 & 20.59 & 18.16 & 15.46 & 14.44 & 14.08 & L6 &    APO \\
SDSSJ1446+0024 & 14:46:00.60 & +00:24:52.0 & 23.39* & 20.74 & 18.54 & 15.56$^{7}$ & 14.59$^{7}$ & 13.80$^{7}$ & L6 (L5)$^{5}$ &    APO$^{4}$ \\
SDSSJ0857+5708 & 08:57:58.45 & +57:08:51.4 & 23.02 & 20.71 & 17.73 & 14.80$^{7}$ & 13.80$^{7}$ & 12.94$^{7}$ & L7 (L8)$^{5}$& APO$^{4}$ \\
SDSSJ0107+0041 & 01:07:52.34 & +00:41:56.1 & 24.06*: & 21.52 & 18.66 & 15.75$^{7}$ & 14.56$^{7}$ & 13.58$^{7}$ & L8 (L5.5)$^{5}$ (L7)$^{6}$&    EDR$^{3}$ \\
SDSSJ1207+0244 & 12:07:47.17 & +02:44:24.8 & 24.37*: & 21.49 & 18.40 & ***** & ***** & ***** & L8 & recent \\
\smallskip
SDSSJ0830+4828 & 08:30:08.12 & +48:28:47.5 & 23.59* & 21.26 & 18.08 & 15.22$^{7}$ & 14.40$^{7}$ & 13.68$^{7}$ & L9$^{5}$& recent$^{4}$ \\
SDSSJ0423$-$0414 & 04:23:48.56 & $-$04:14:03.5 & 22.64 & 20.21 & 17.33 & 14.45 & 13.44 & 12.94 & TO$^{5}$ (L5)$^{6}$ &    APO$^{3,8}$ \\
SDSSJ0151+1244 & 01:51:41.68 & +12:44:29.6 & 24.31*: & 22.84*: & 19.45 & 16.25$^{7}$ & 15.54$^{7}$ & 15.18$^{7}$ & T1$^{5}$& APO$^{4}$ \\
SDSSJ1254$-$0122 & 12:54:53.90 & $-$01:22:47.4 & 24.06*: & 22.26*: & 18.01 & 14.88 & 14.04 & 13.83 & T2$^{5}$ & recent$^{2}$ \\
SDSSJ1021$-$0304 & 10:21:09.69 & $-$03:04:20.1 & 23.31* & 22.39* & 19.31 & 16.26 & 15.33 & 15.10 & T3$^{5}$ &    APO$^{2}$ \\
\enddata
\tablecomments{
Comments: \\
$*$ --- Magnitude is fainter than formal 5$\sigma$ limits: $r^*=23.1; i^*=22.3; z^*=20.8$ \\
: --- Magnitude uncertainty is greater than $r^*=0.5; i^*=0.2; z^*=0.2$ \\
1 --- First identified by Fan et al. (2000)\\
2 --- First identified by Leggett et al. (2000)\\
3 --- First identified by Schneider et al. (2002)\\
4 --- First identified by Geballe et al. (2002)\\   
5 --- Spectral type given by Geballe et al. (2002)\\
6 --- Spectral type given by Schneider et al. (2002)\\
7 --- near-IR observations reported by Leggett et al. (2002), in the MKO filter system\\
8 --- Photometry from preliminary processing of data not contained in the final
SDSS database\\
9 --- Identified by, and spectral type from, Gizis et al. (2000) as 2MASSW J1707+6439\\
10 -- Identified by, and spectral type from, Martin et al. (1999) as DENISp 1159.6+0057\\
11 -- Position within released 2MASS survey limits; object not detected in IR\\
12 -- The K band data are from 2MASS unless otherwise noted, and were obtained
in the 2MASS K$_s$ filter \\
13 -- Identified by, and spectral type from, Kirkpatrick et al. (2000) as
2MASSW J0801+4628 \\
}
\end{deluxetable}

\clearpage

% This is where the tables go.
%in the text refer to it as:
%Templates are listed in Table~\ref{table-templates}.

\begin{figure}
\figurenum{5a}
\epsscale{0.95}
\plotone{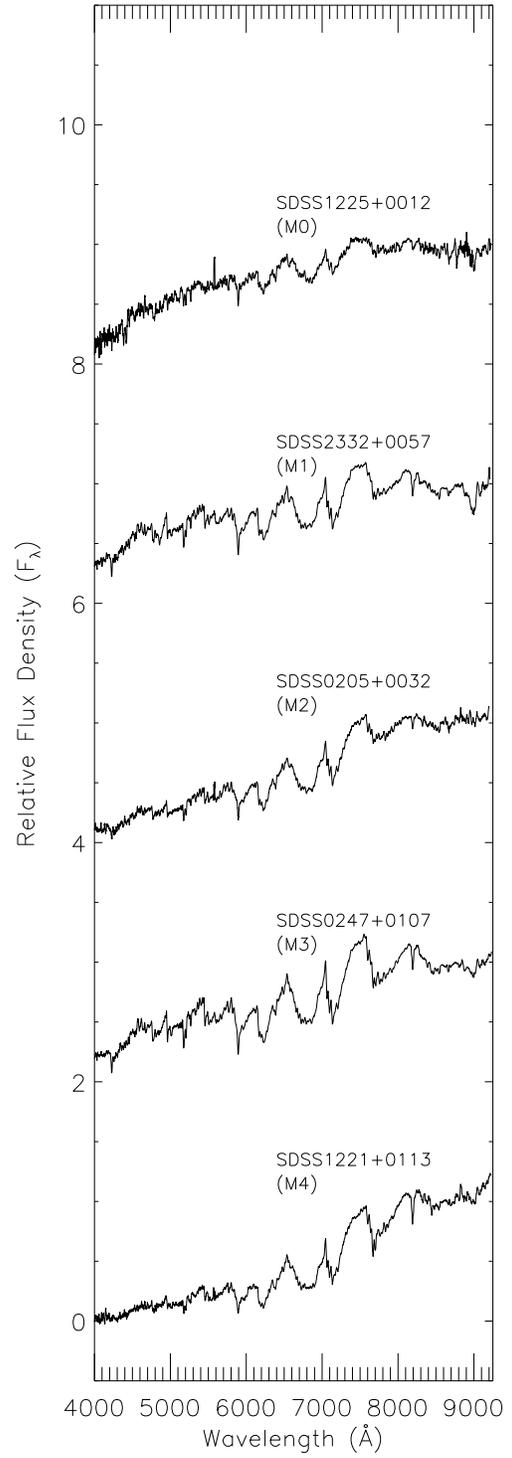}
\caption{The SDSS spectrum with highest signal-to-noise ratio
at each spectral type is shown.  These spectra comprise a set 
of templates to be used in the SDSS pipeline processing for 
automatic spectral type identification in the future.}
\label{fig-sdsstemp1}
\end{figure}

\begin{figure}
\figurenum{5b}
\epsscale{0.95}
\plotone{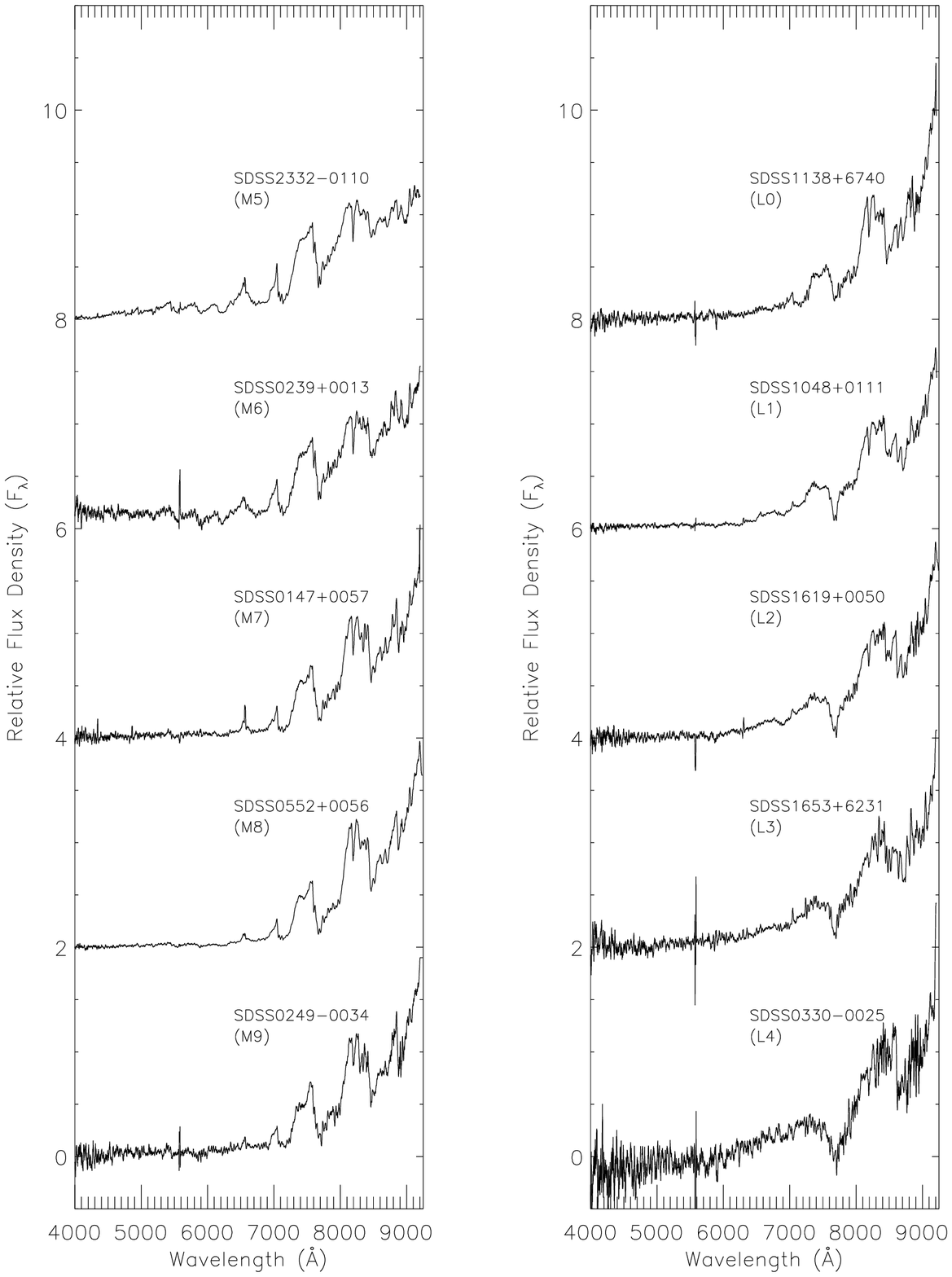}
\caption{Same as Figure~\ref{fig-sdsstemp1}.}
\label{fig-sdsstemp2}
\end{figure}

\begin{figure}
\figurenum{6a}
\epsscale{0.95}
\plotone{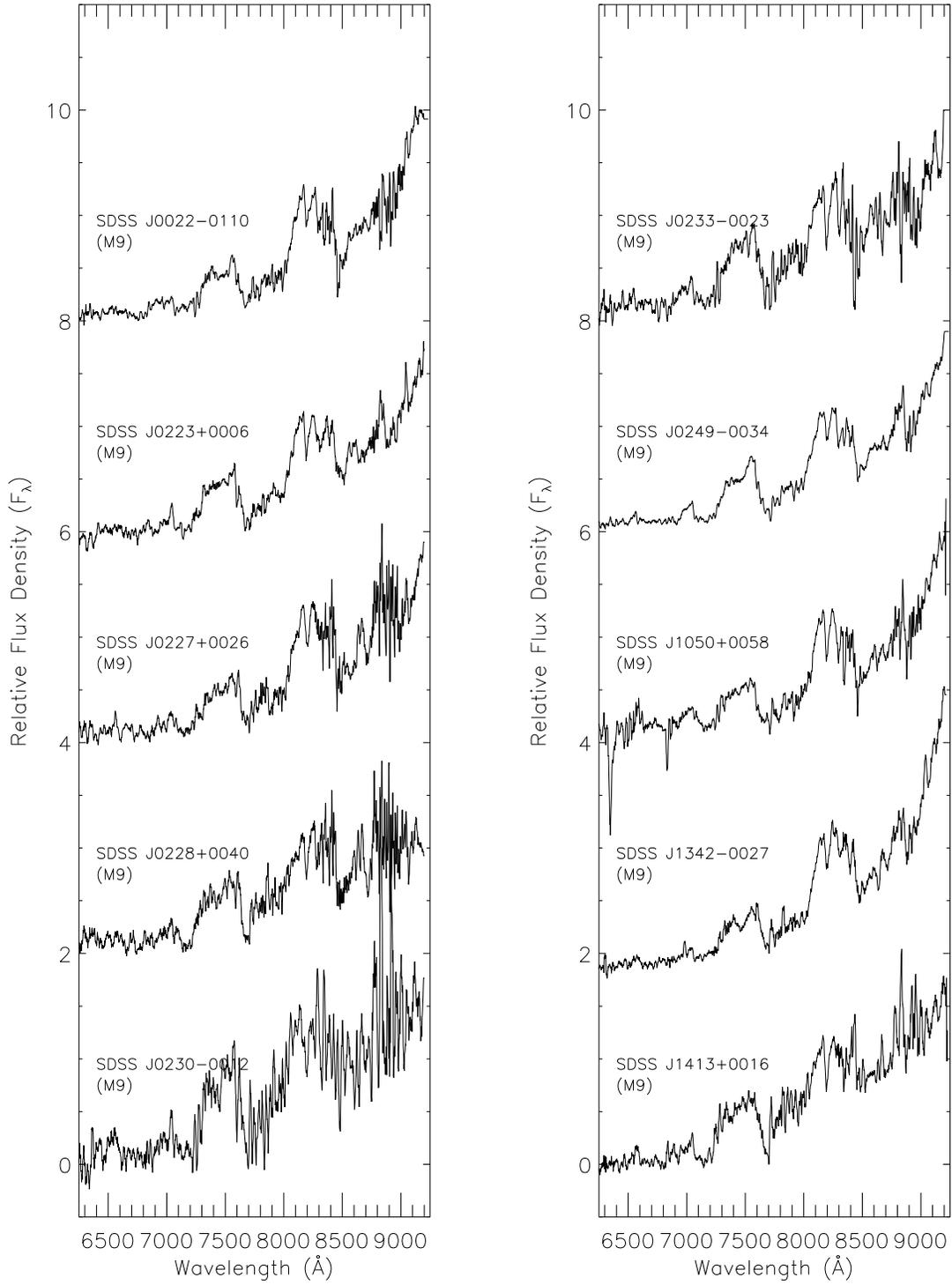}
\caption{SDSS spectra for all objects M9 and later are shown.
The spectral types are ours, as given in Table~\ref{table-finaldata},
except for the L9 and T dwarfs where they are taken from previously
published sources as described in the table notes.}
\label{fig-sdssspec1}
\end{figure}

\begin{figure}
\figurenum{6b}
\epsscale{0.95}
\plotone{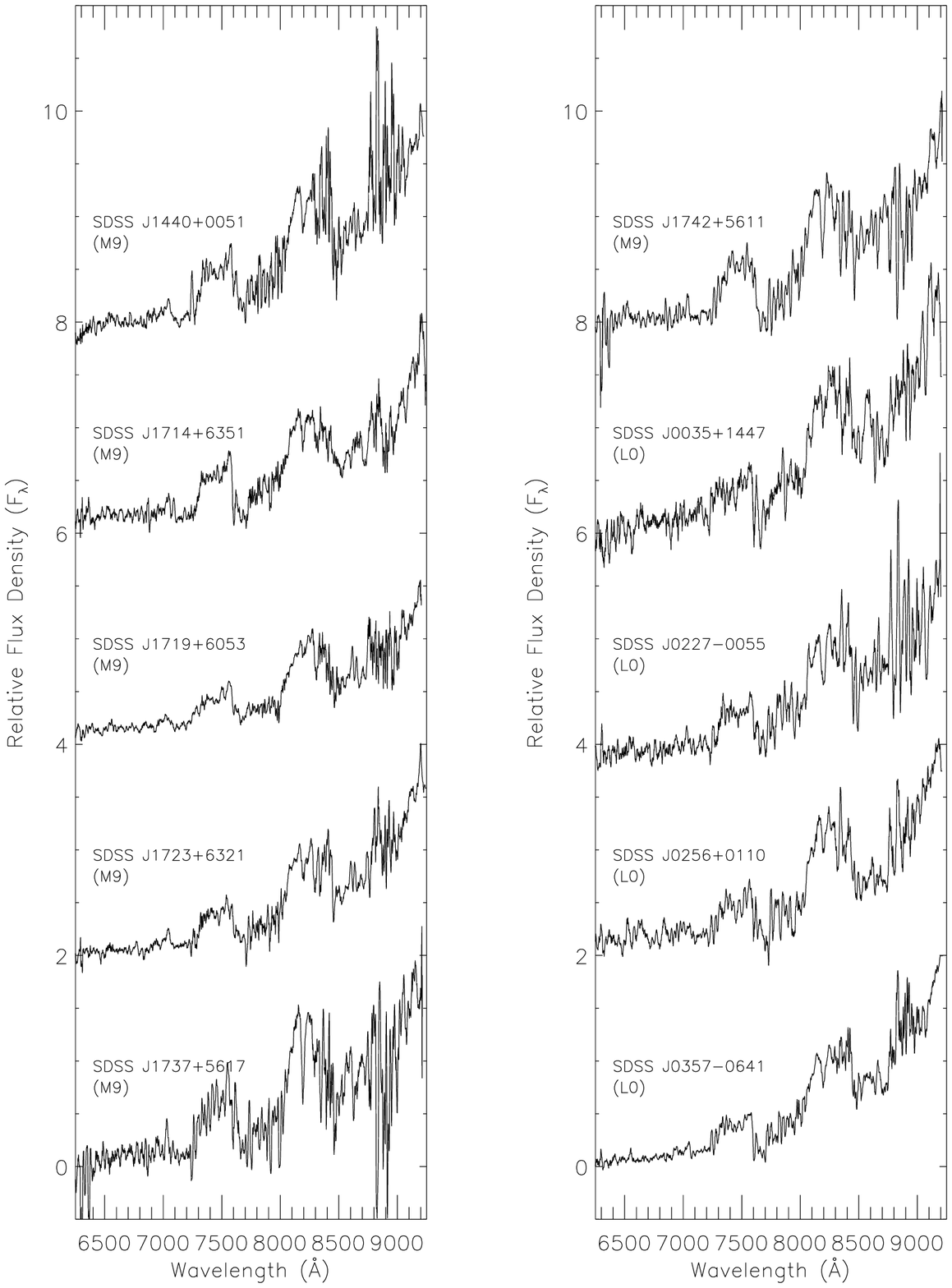}
\caption{Same as Figure~\ref{fig-sdssspec1}.}
\label{fig-sdssspec2}
\end{figure}

\begin{figure}
\figurenum{6c}
\epsscale{0.95}
\plotone{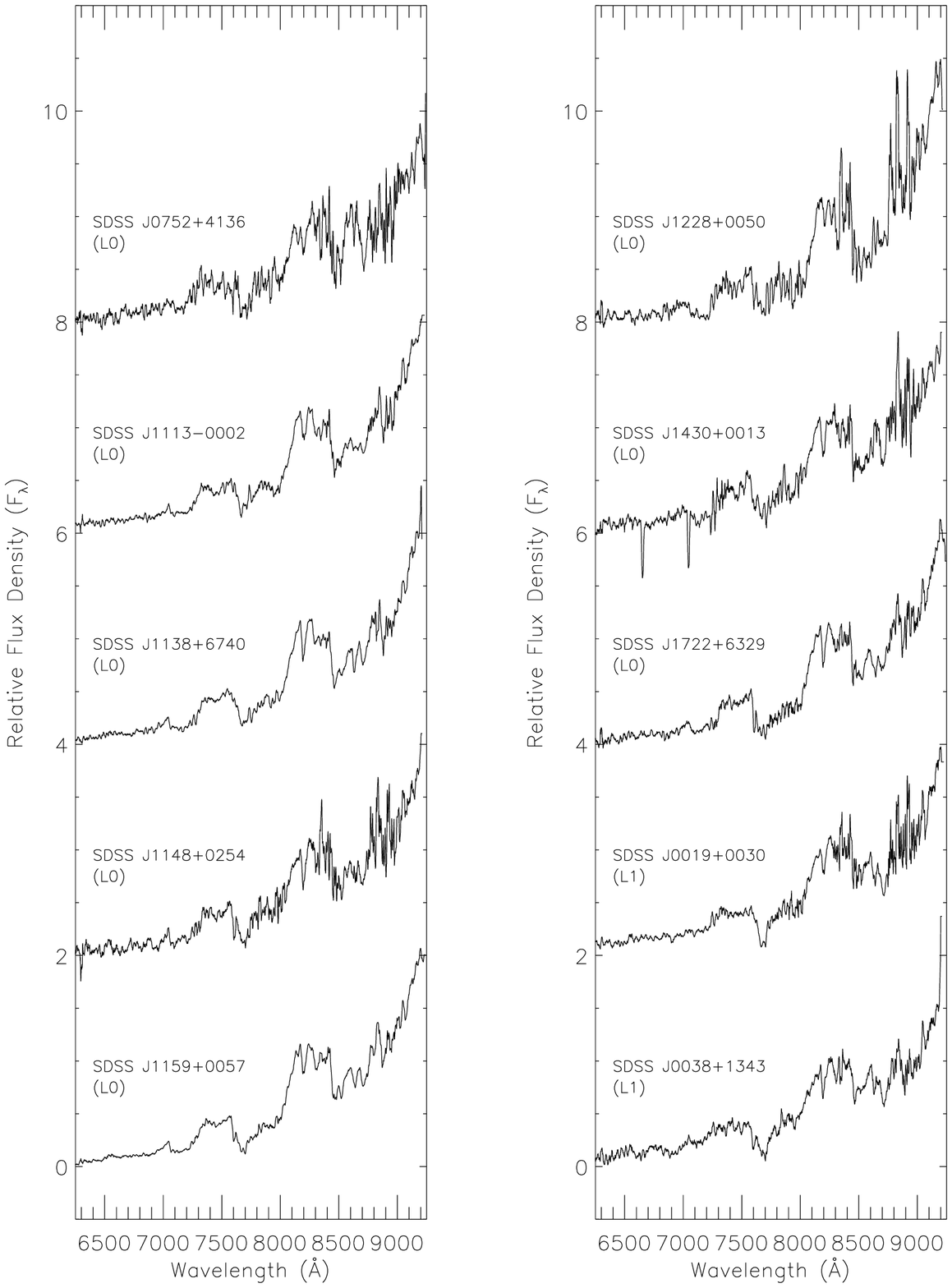}
\caption{Same as Figure~\ref{fig-sdssspec1}.}
\label{fig-sdssspec3}
\end{figure}

\begin{figure}
\figurenum{6d}
\epsscale{0.95}
\plotone{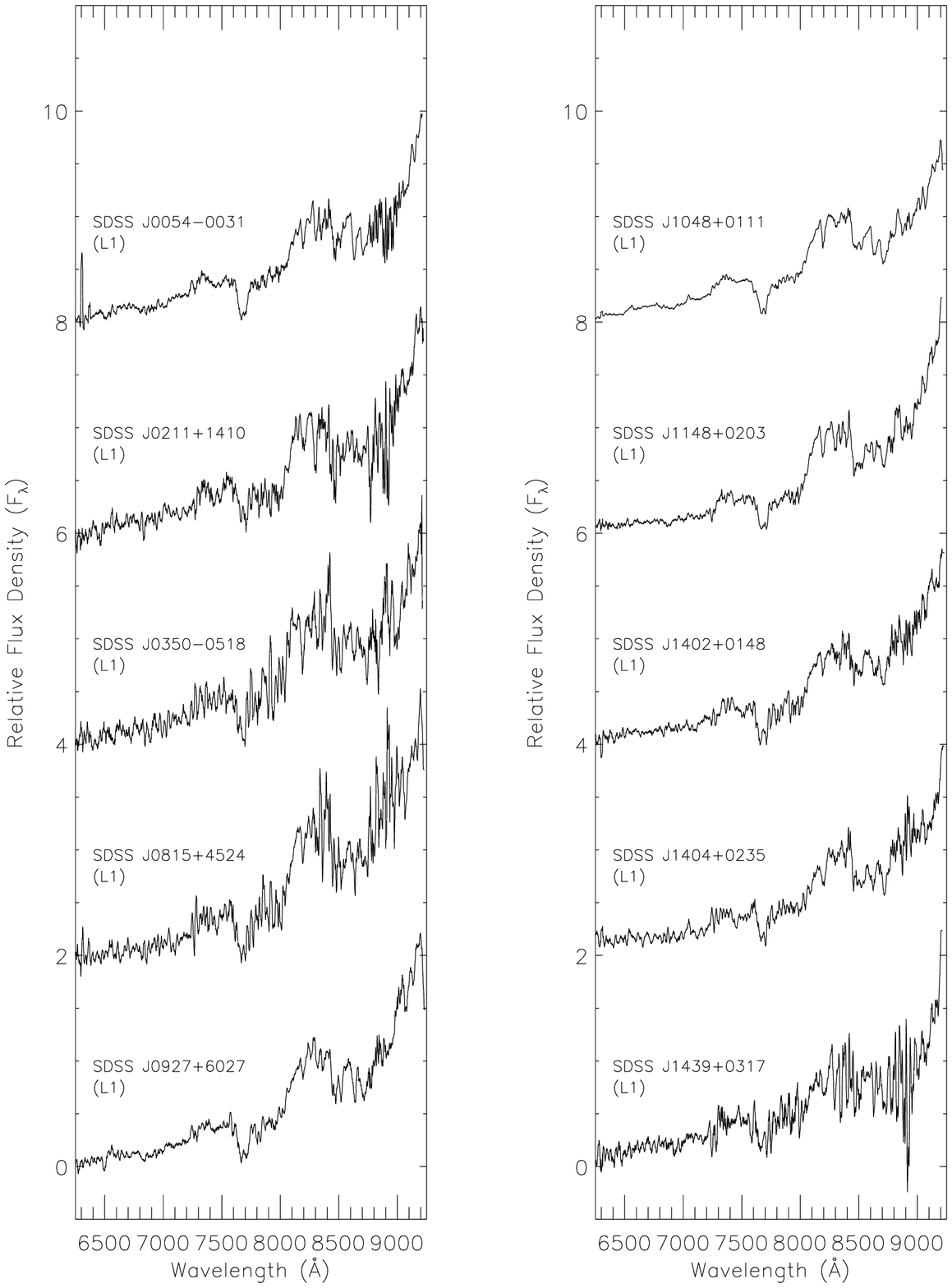}
\caption{Same as Figure~\ref{fig-sdssspec1}.}
\label{fig-sdssspec4}
\end{figure}

\begin{figure}
\figurenum{6e}
\epsscale{0.95}
\plotone{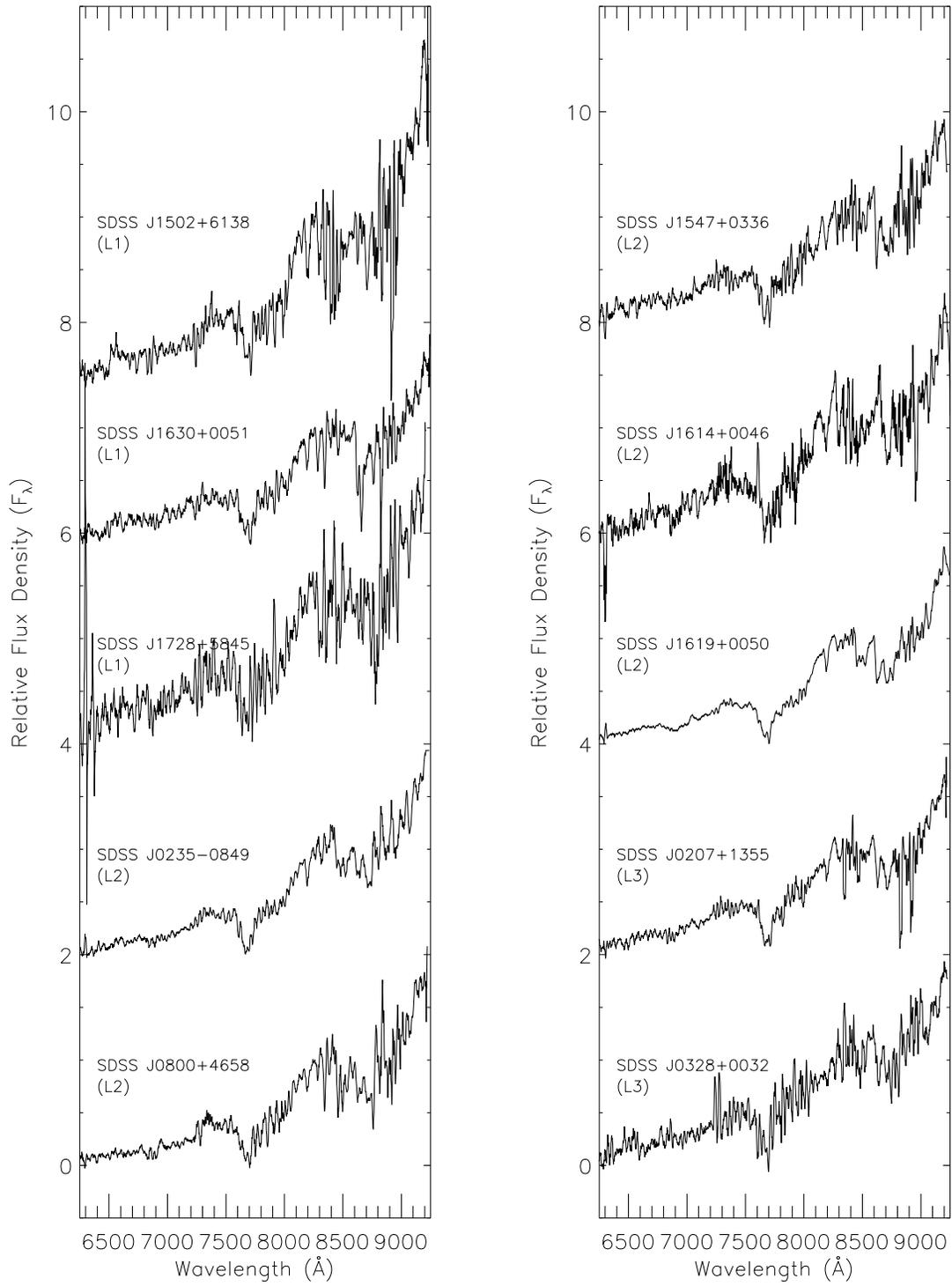}
\caption{Same as Figure~\ref{fig-sdssspec1}.  Note
that we have not attempted to remove noise spikes
such as those in the left panel near 6300\AA.}
\label{fig-sdssspec5}
\end{figure}

\begin{figure}
\figurenum{6f}
\epsscale{0.95}
\plotone{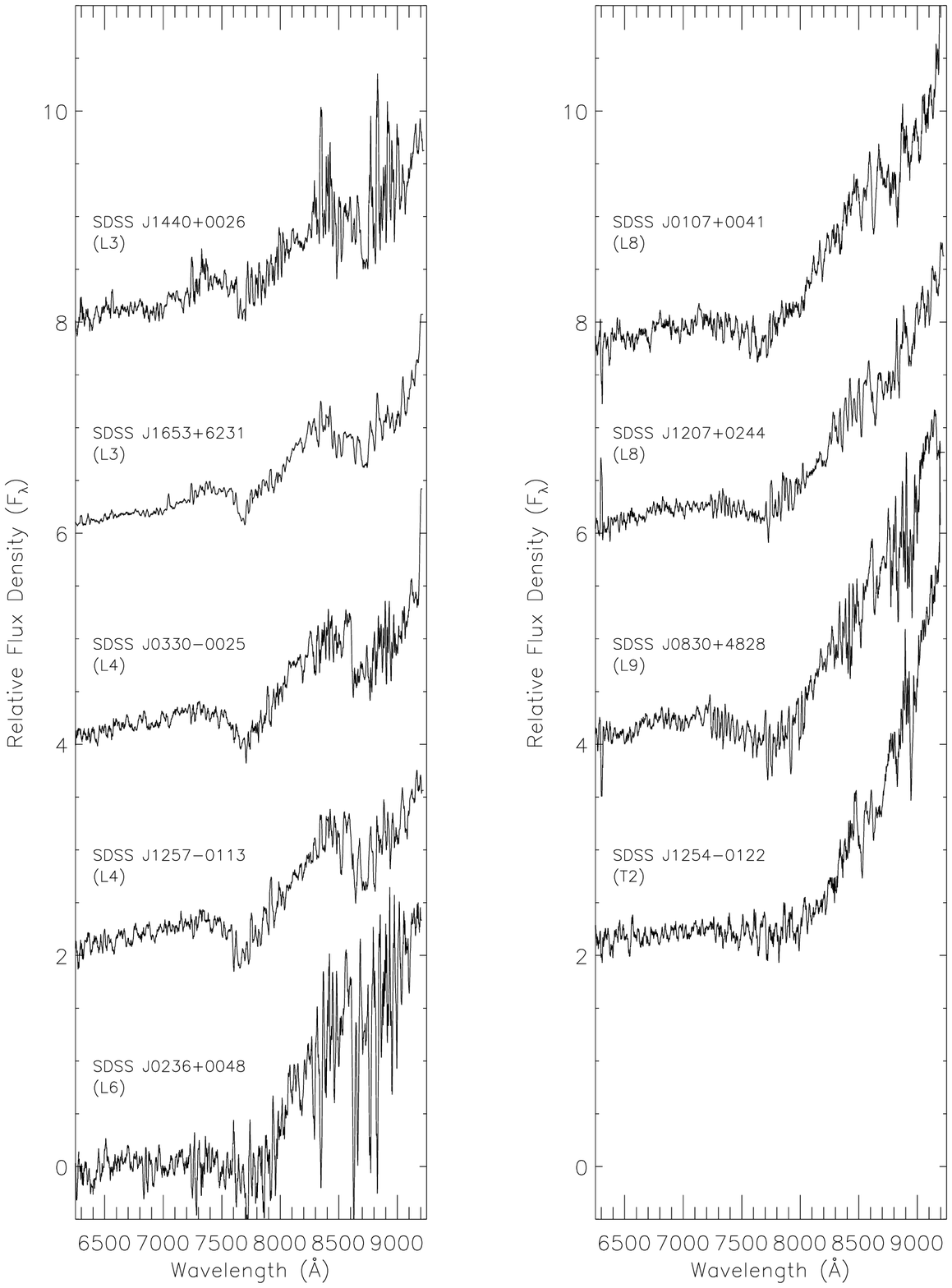}
\caption{Same as Figure~\ref{fig-sdssspec1}.}
\label{fig-sdssspec6}
\end{figure}

\begin{figure}
\figurenum{7a}
\epsscale{0.95}
\plotone{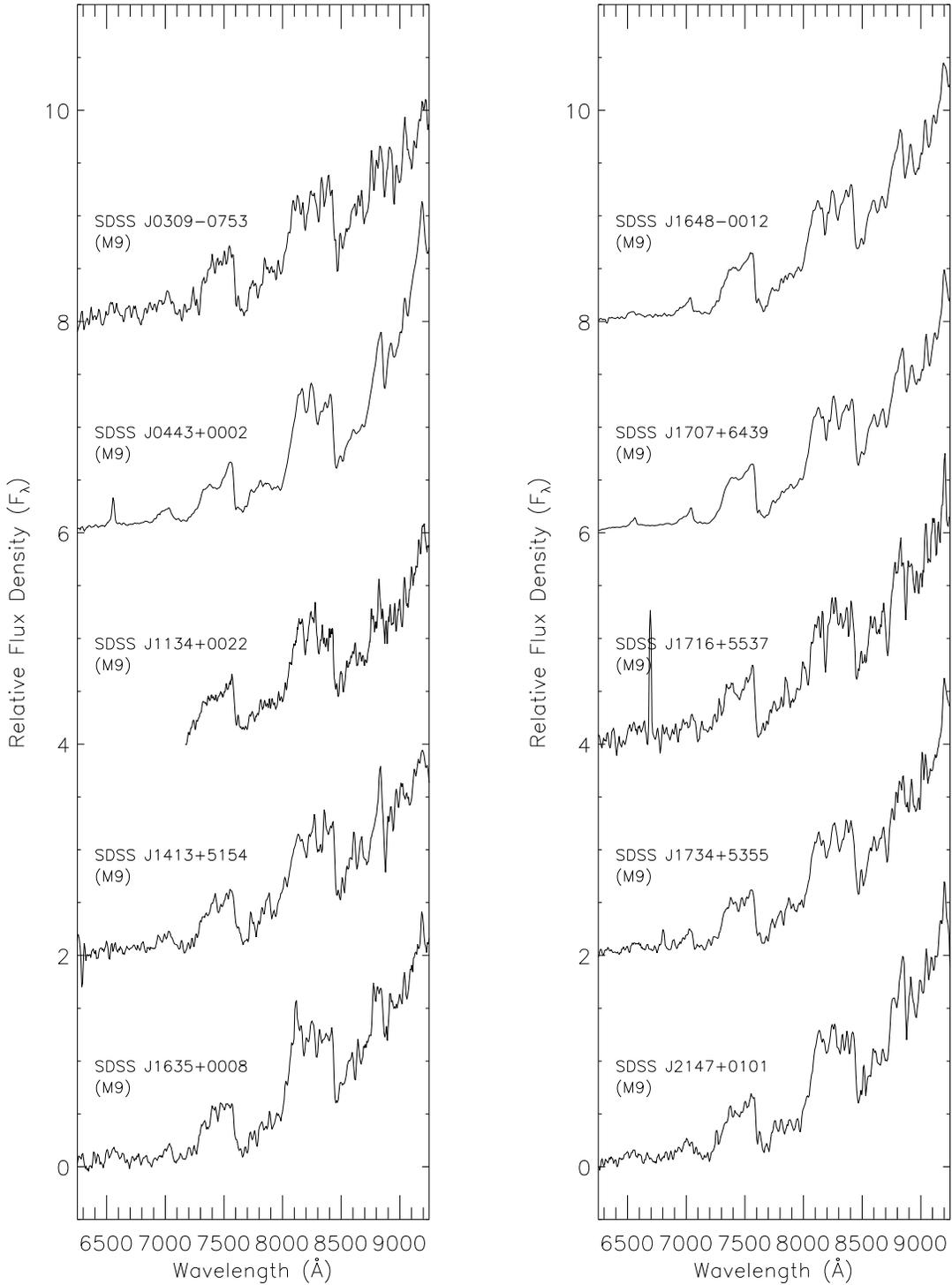}
\caption{APO 3.5m spectra for all objects M9 and later are shown.
The spectral types are ours, as given in Table~\ref{table-finaldata},
except for the T dwarfs where they are taken from previously
published sources as described in the table notes.
These targets were chosen from SDSS imaging data to be very red
in the SDSS $(i^*-z^*)$ color, producing most of the new mid-late 
L dwarf spectra in our sample.  Note that we have not attempted
to remove noise spikes such as the one near 6700\AA\ in SDSSJ1716+5537.}
\label{fig-apospec1}
\end{figure}

\begin{figure}
\figurenum{7b}
\epsscale{0.95}
\plotone{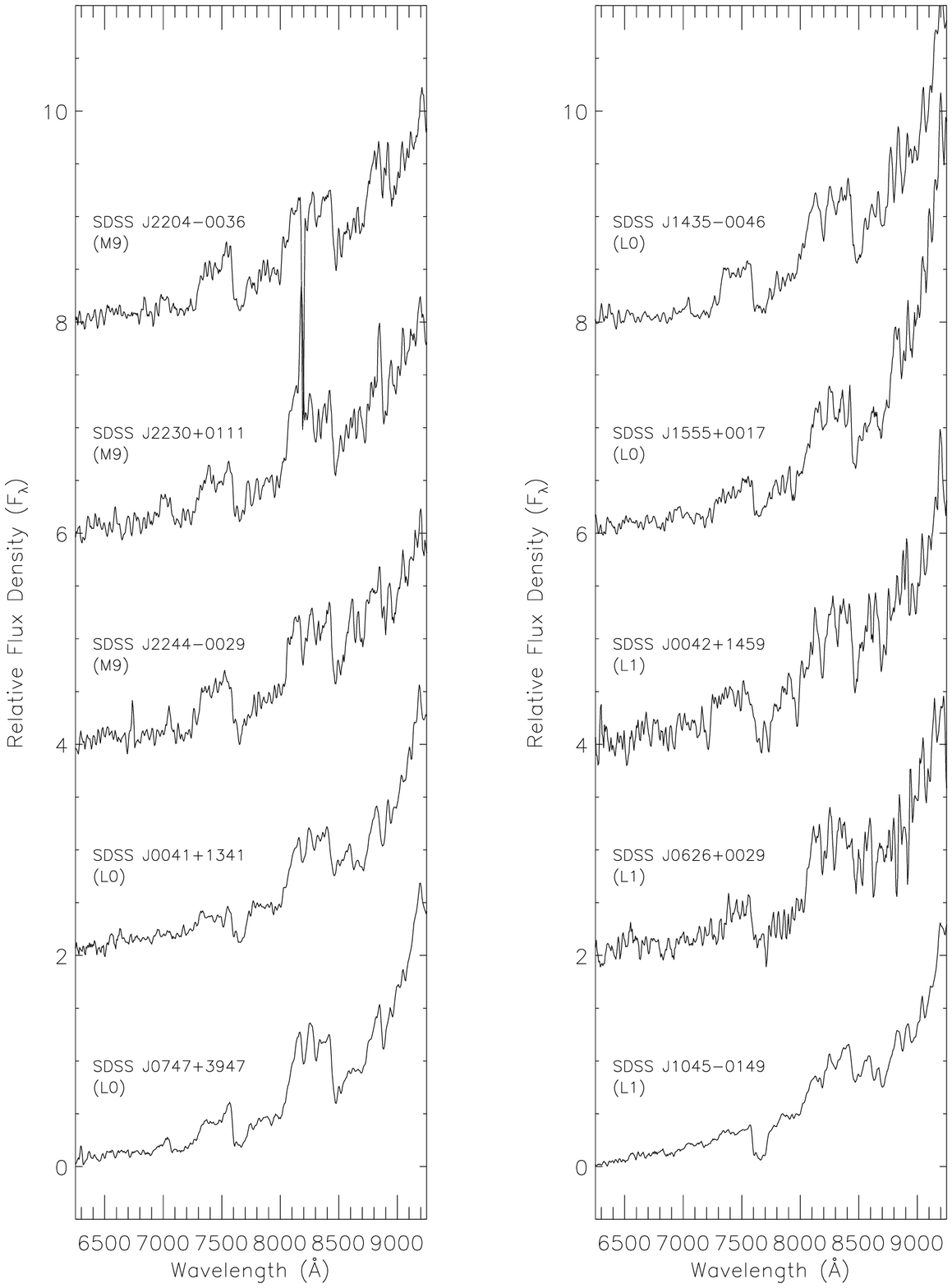}
\caption{Same as Figure~\ref{fig-apospec1}.}
\label{fig-apospec2}
\end{figure}

\begin{figure}
\figurenum{7c}
\epsscale{0.95}
\plotone{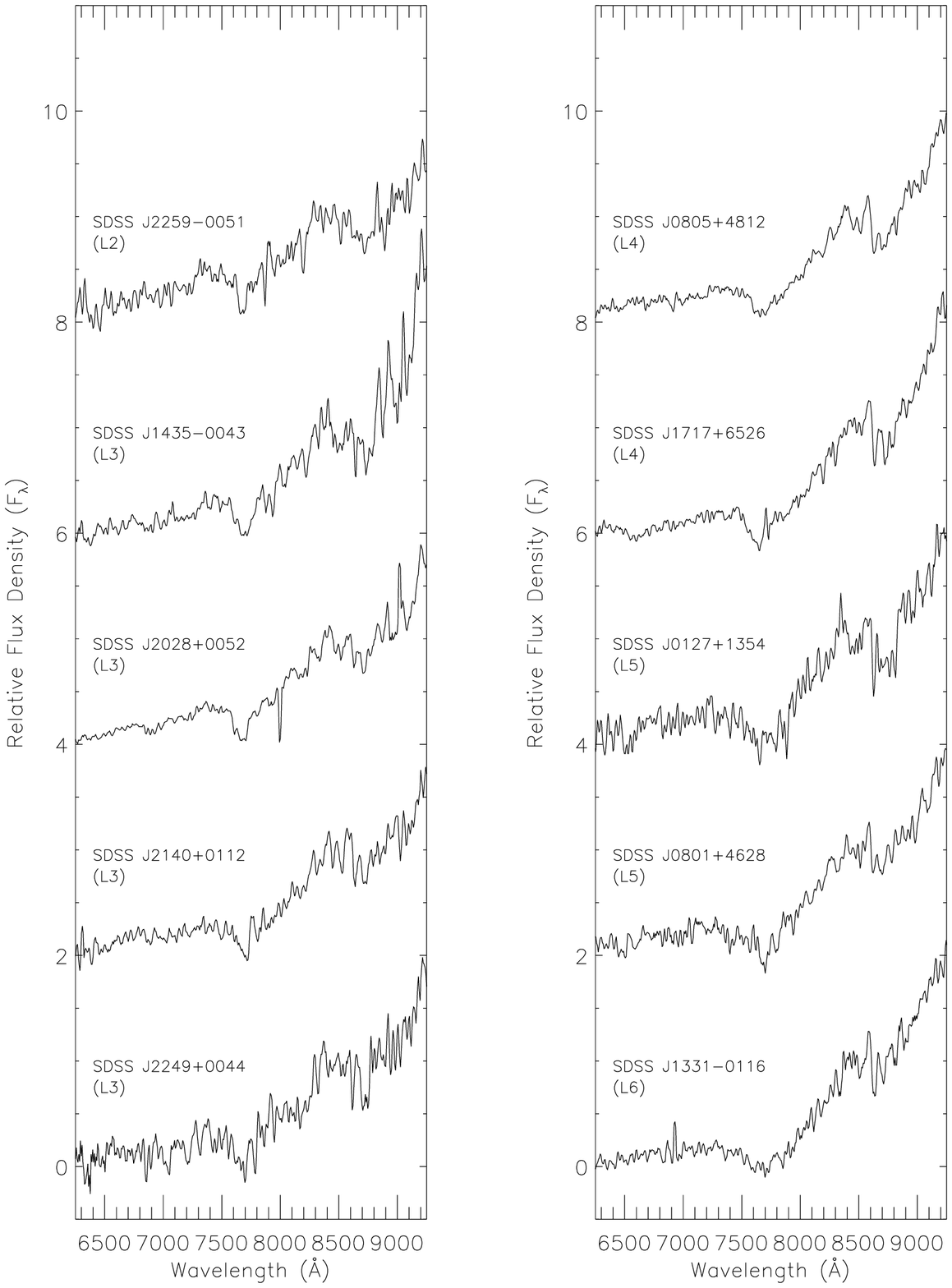}
\caption{Same as Figure~\ref{fig-apospec1}.}
\label{fig-apospec3}
\end{figure}

\begin{figure}
\figurenum{7d}
\epsscale{0.95}
\plotone{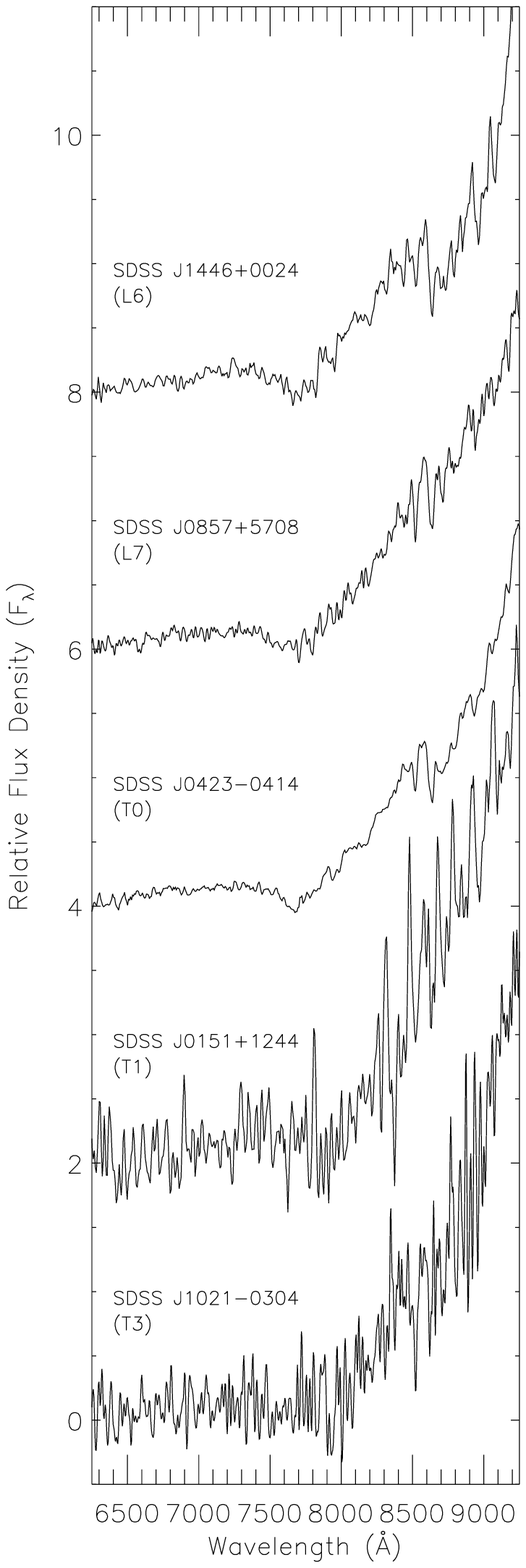}
\caption{Same as Figure~\ref{fig-apospec1}.}
\label{fig-apospec4}
\end{figure}

\begin{figure}
\figurenum{8}
\epsscale{0.95}
\plotone{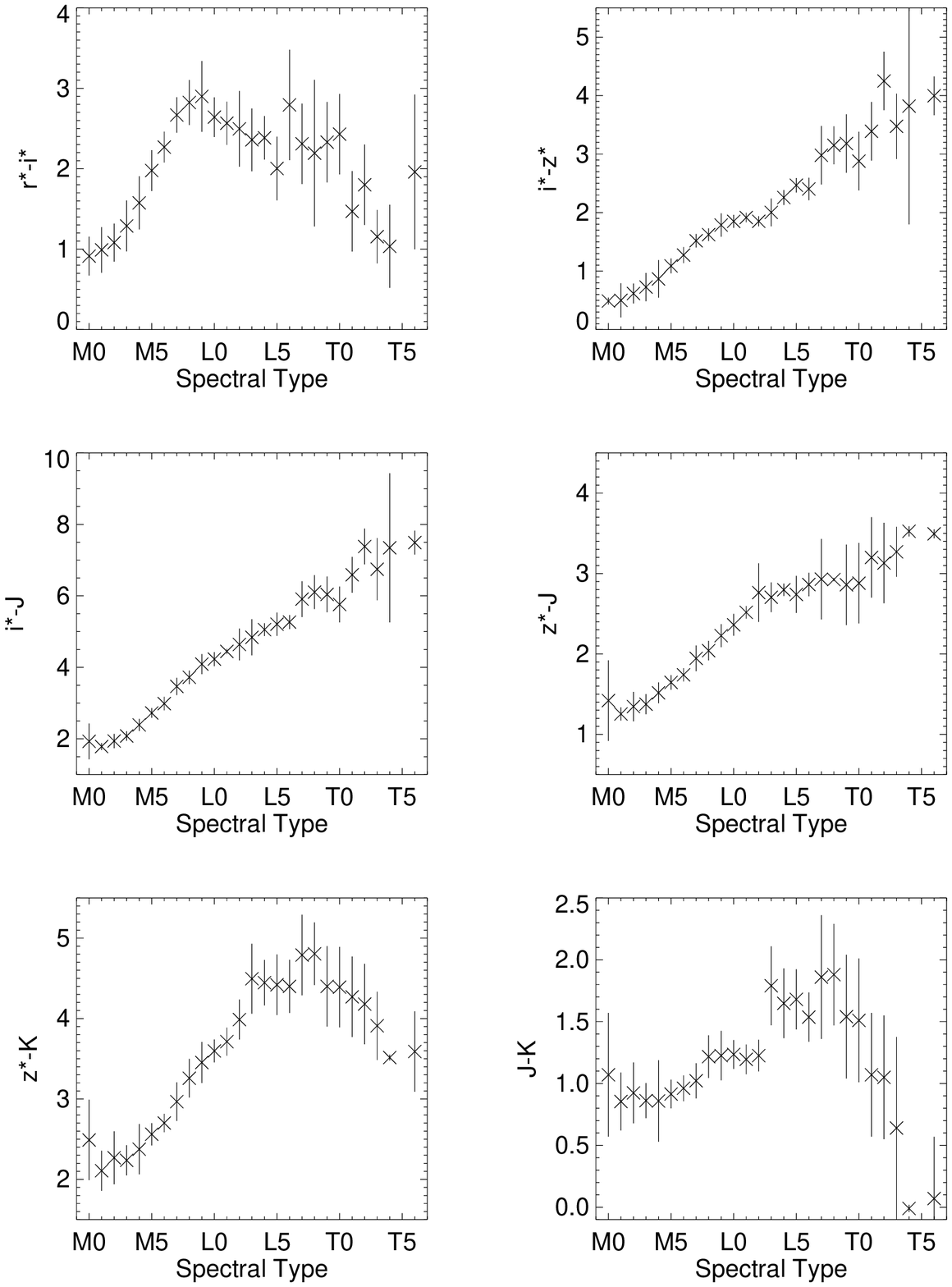}
\caption{Average color -- spectral type relations for the full sample
are shown.  The error bars indicate the standard deviation of
the mean color at each spectral type.  If only one object of a given
type was measured, the standard deviation was set to 0.5.}
\label{fig-avesptcolor}
\end{figure}

\begin{figure}
\figurenum{9}
\epsscale{0.95}
\plotone{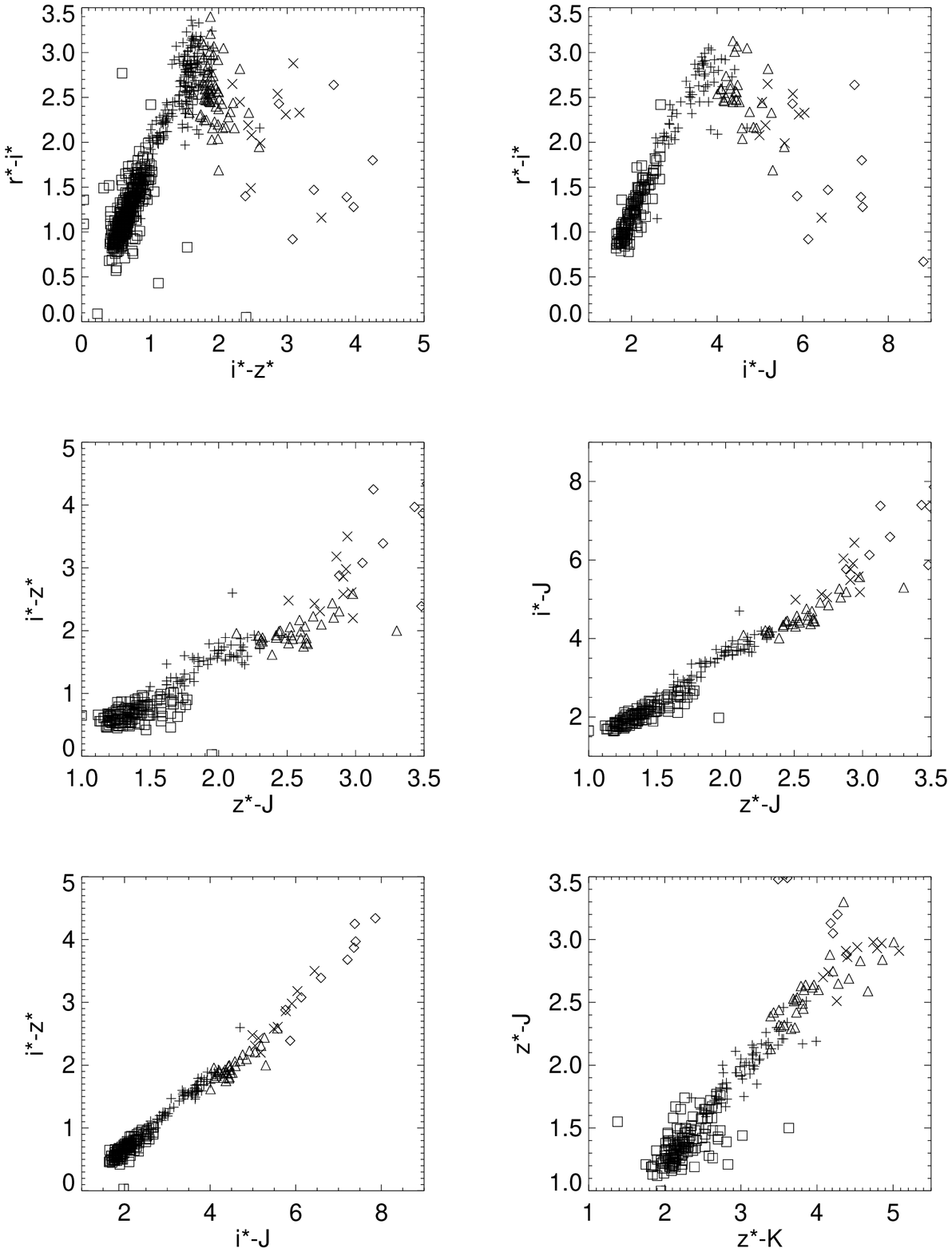}
\caption{Color-color relations for the full sample
are shown.  Open squares have types M0-M4; plus signs have types
M5-M9; open triangles have types L0-L4; crosses have types L5-L9;
and open diamonds have types T0-T8.}
\label{fig-colorcolor}
\end{figure}

\clearpage

\begin{figure}
\figurenum{10}
\epsscale{0.95}
\plotone{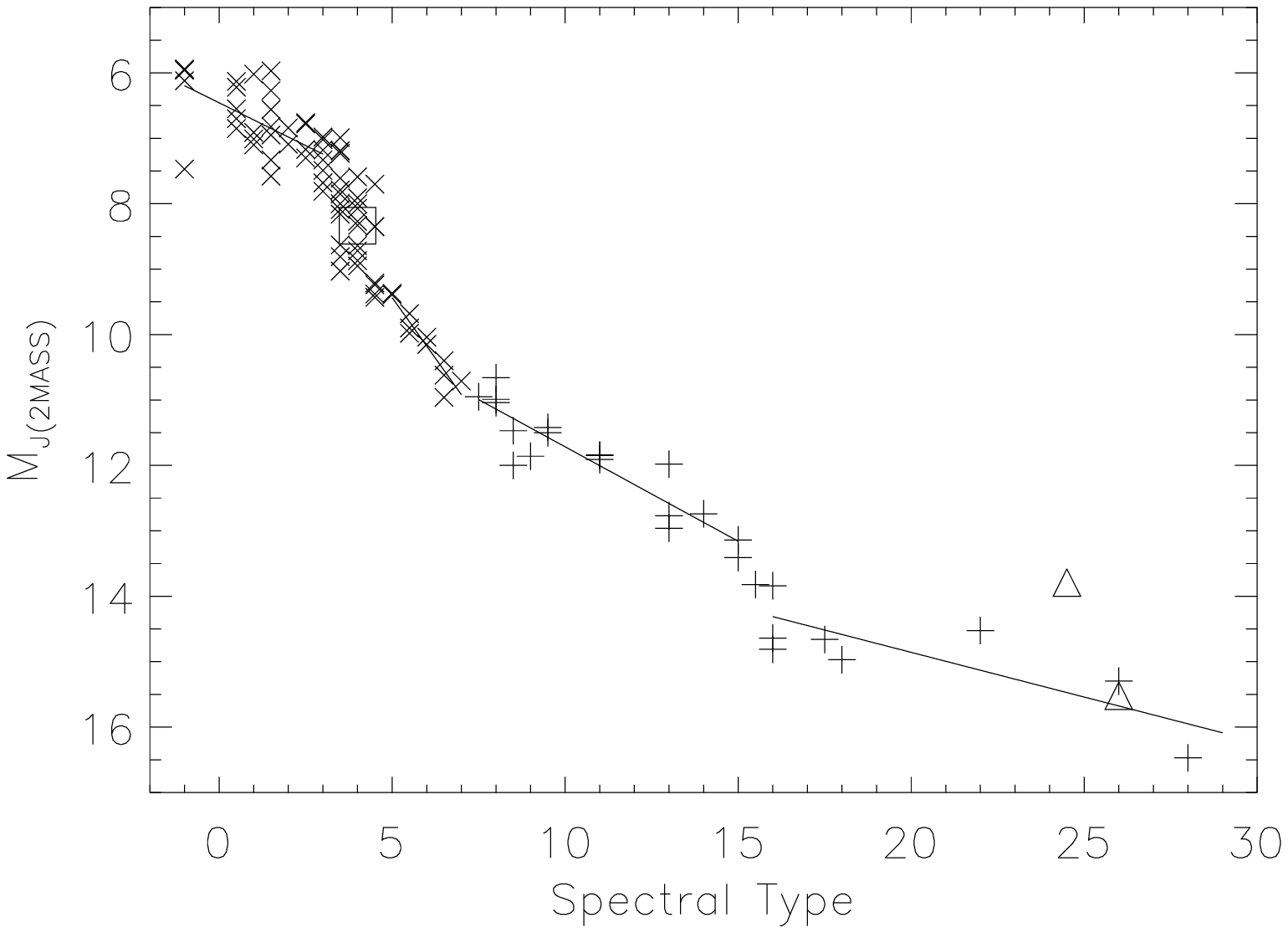}
\caption{Calibrated M$_J$ vs spectral type relation for
late type dwarfs with measured 2MASS J magnitudes and distances
from trigonometric parallax determinations. Crosses (types M0-M7)
come from the 8-pc sample; plus signs (types M8-T8) 
are from several sources as described in 
the text.  The open triangles are 2MASSJ0559-1404 (T4) and Gl 299B (T6)
which are not included in the fit (see text).  The solid lines 
represent our best piecewise linear fit to the data.  The large
open square at type M4 is our adopted mean value.}
\label{fig-specpical}
\end{figure}

\begin{figure}
\figurenum{11}
\epsscale{0.95}
\plotone{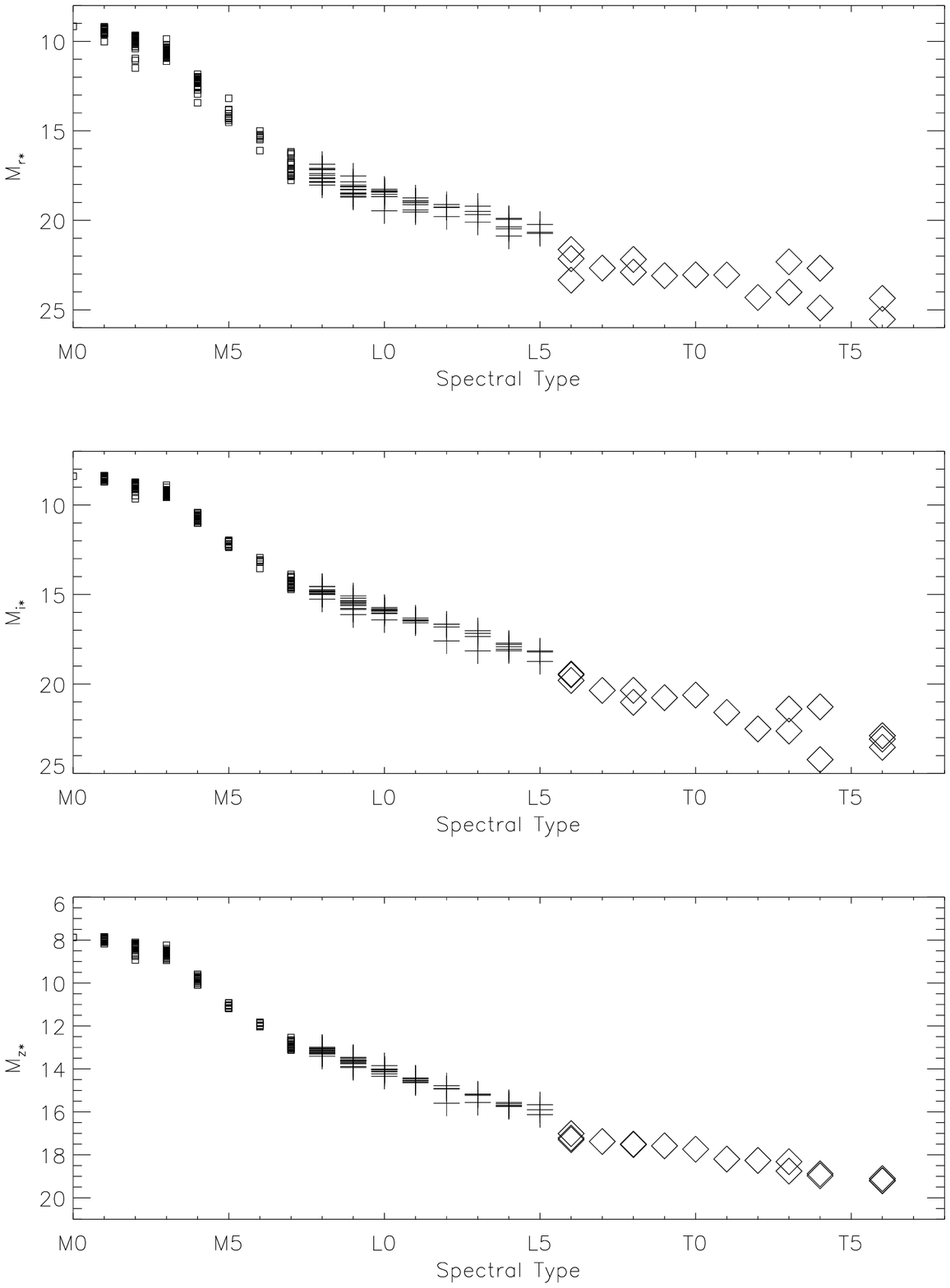}
\caption{Spectroscopic parallax relations for 
M$_{r^*}$, M$_{i^*}$ and M$_{z^*}$, using the fits
shown in Figure~\ref{fig-specpical} and our measured
colors.  Only M dwarfs with 2MASS J magnitudes are shown.
L and T dwarfs with MKO J magnitudes were transformed using
Equation~\ref{equation-mko2mdif}.  Small open squares depict 
types M0-M7, plus signs are used for M8-L5 
and open diamonds for L6-T8.}
\label{fig-specpi3}
\end{figure}

\begin{figure}
\figurenum{12}
\epsscale{0.95}
\plotone{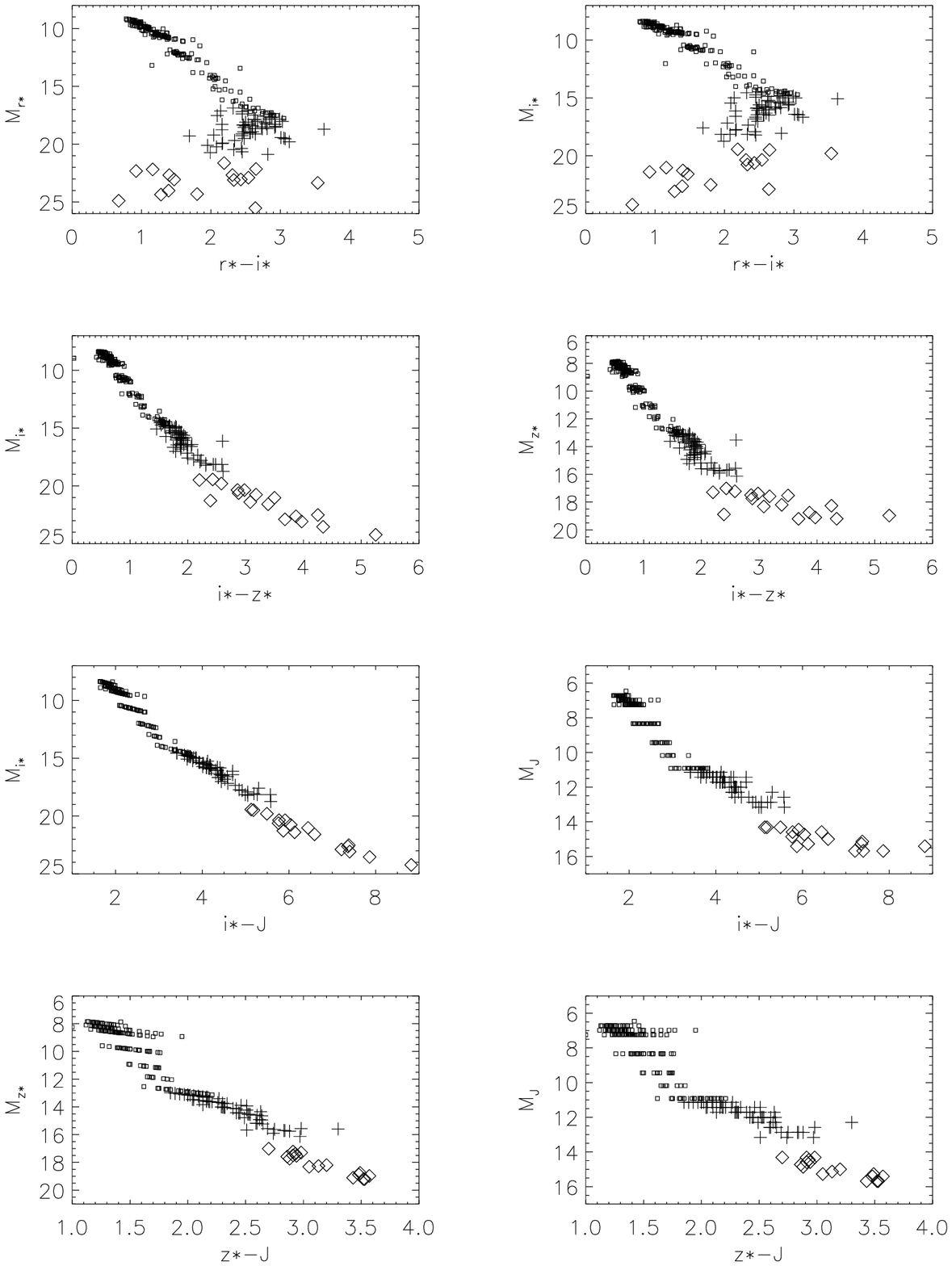}
\caption{Photometric parallax relations for color
combinations using the $r^*,i^*,z^*$ and 2MASS J magnitudes.
Symbols are the same as in Figure~\ref{fig-specpi3}.}
\label{fig-photpiall}
\end{figure}

\begin{figure}
\figurenum{13}
\epsscale{0.95}
\plotone{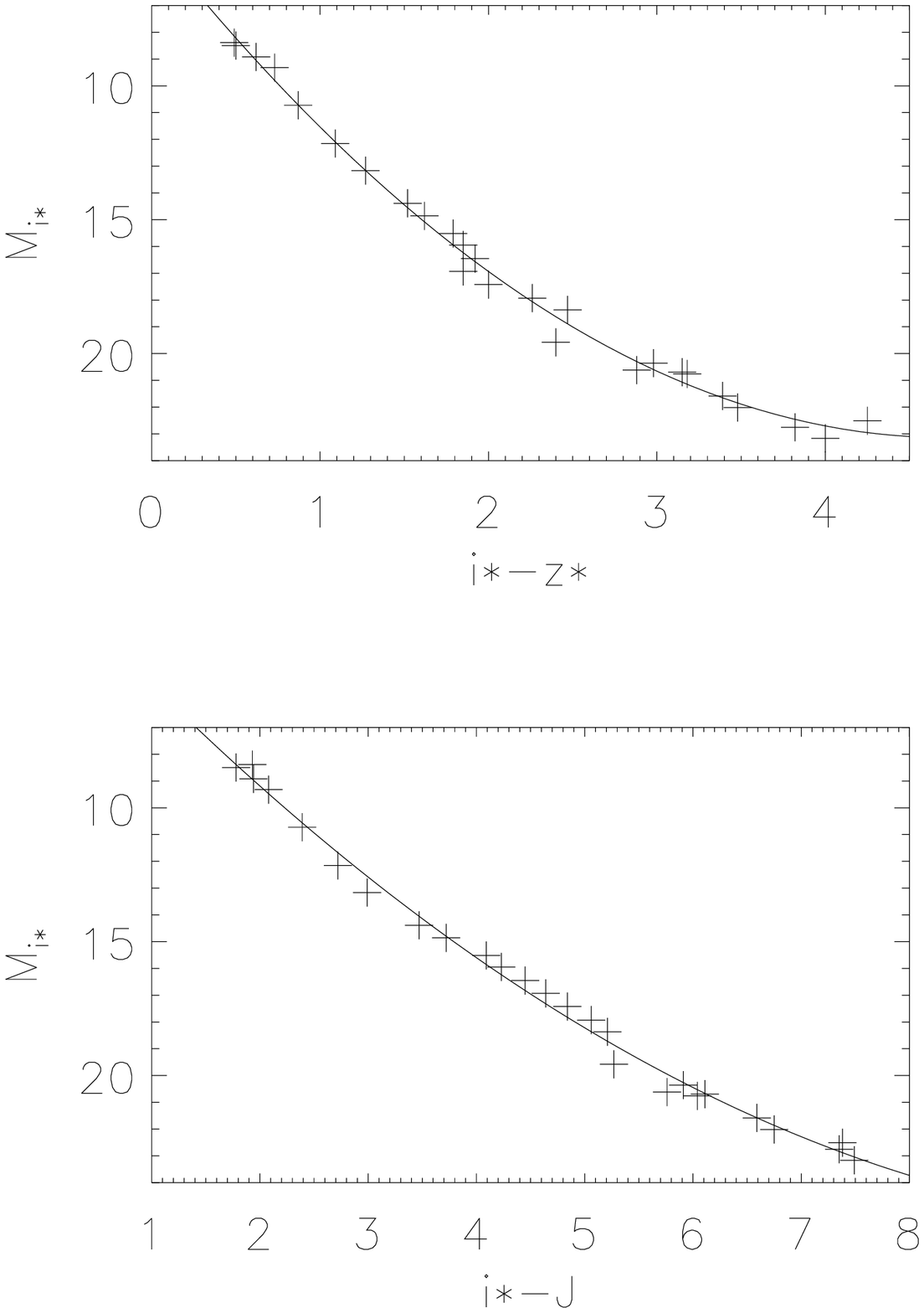}
\caption{Adopted photometric parallax relations
for M$_{i^*}$,$(i^*-z^*)$ and M$_{i^*}$,$(i^*-$J).
The best fit second-order polynomial relations are shown
by solid lines and given in the text.}
\label{fig-absmags}
\end{figure}

\begin{figure}
\figurenum{14}
\epsscale{0.95}
\plotone{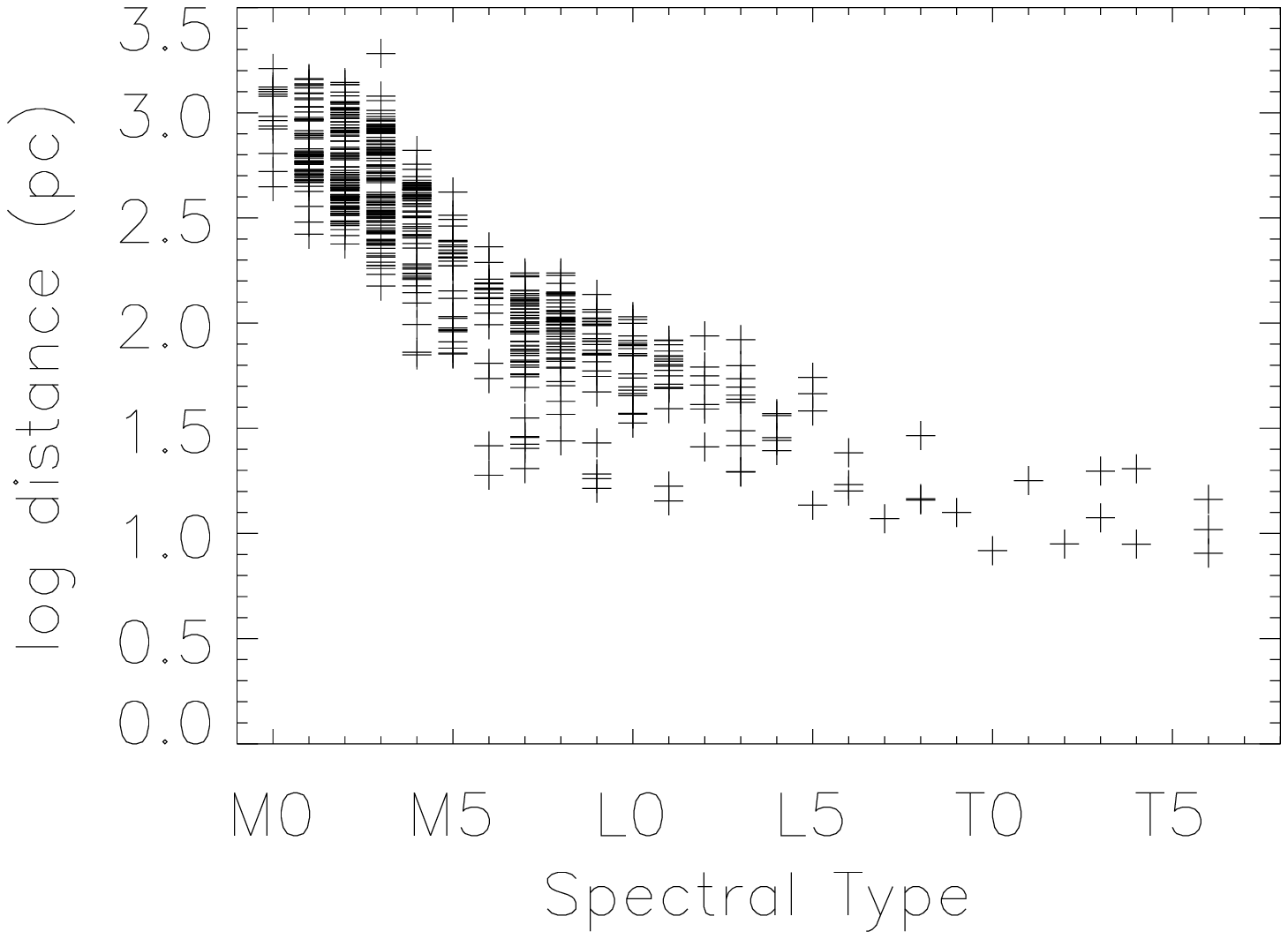}
\caption{Spectroscopic distance estimates for all objects with measured
spectral types.  The early M dwarfs extend out to $\sim$ 1.5 kpc
while the L dwarfs are confined to $\sim$ 100 pc and
the T dwarfs to $\sim$ 20 pc.}
\label{fig-distlog}
\end{figure}

%\clearpage

\end{document}